%% file: main.tex
\pdfoutput=1
\documentclass[acmsmall,screen]{acmart}

\setcopyright{none}
\acmJournal{TRETS}

\usepackage{xspace}
\usepackage{enumitem}
\usepackage{algorithm}
\usepackage{algpseudocode}

\newcommand{\sysname}{Cruella\xspace}

\renewcommand\footnotetextcopyrightpermission[1]{}
\makeatletter
\AtBeginDocument{%
  \g@addto@macro\ps@standardpagestyle{\fancyfoot{}}%
  \g@addto@macro\ps@firstpagestyle{\fancyfoot{}}%
  \pagestyle{standardpagestyle}}
\makeatother

\begin{document}

\title{Cruella: A Unified Microscaling and Floating-Point Dot-Product Unit for LLM Inference on FPGAs}

\author{Jinho Yang}
\orcid{0009-0002-6241-5650}
\email{jinho.yang@hyperaccel.ai}
\affiliation{\institution{HyperAccel}\city{Seoul}\country{South Korea}}

\author{Taehyeong Kim}
\orcid{0009-0000-8663-1521}
\email{taehyeong.kim@hyperaccel.ai}
\affiliation{\institution{HyperAccel}\city{Seoul}\country{South Korea}}

\author{Mincheol Cha}
\orcid{0009-0003-1359-2439}
\email{mincheol.cha@hyperaccel.ai}
\affiliation{\institution{HyperAccel}\city{Seoul}\country{South Korea}}

\author{Hyeoncheol Kim}
\orcid{0009-0009-6819-2591}
\email{hyeoncheol.kim@hyperaccel.ai}
\affiliation{\institution{HyperAccel}\city{Seoul}\country{South Korea}}

\author{Donggeun Kim}
\orcid{0009-0002-2544-564X}
\email{donggeun.kim@hyperaccel.ai}
\affiliation{\institution{HyperAccel}\city{Seoul}\country{South Korea}}

\author{Nahye Kim}
\orcid{0009-0000-7439-0257}
\email{iachristian@kaist.ac.kr}
\affiliation{\institution{KAIST}\city{Daejeon}\country{South Korea}}

\author{Yujin Lee}
\orcid{0009-0007-2545-6046}
\email{YUJIN004@e.ntu.edu.sg}
\affiliation{\institution{Nanyang Technological University}\city{Singapore}\country{Singapore}}

\author{Ryeowook Ko}
\orcid{0009-0006-3829-7761}
\email{ryeowook.ko@hyperaccel.ai}
\affiliation{\institution{HyperAccel}\city{Seoul}\country{South Korea}}

\author{Woong Gyu Park}
\orcid{0009-0002-4106-8039}
\email{woonggyu.park@hyperaccel.ai}
\affiliation{\institution{HyperAccel}\city{Seoul}\country{South Korea}}

\author{Jiwoong Park}
\orcid{0009-0009-7813-2891}
\email{jiwoong.park@hyperaccel.ai}
\affiliation{\institution{HyperAccel}\city{Seoul}\country{South Korea}}

\author{Junsoo Kim}
\orcid{0000-0001-6680-2602}
\email{js.kim@hyperaccel.ai}
\affiliation{\institution{HyperAccel}\city{Seoul}\country{South Korea}}

\author{Tae-Hyoung Kim}
\orcid{0000-0002-1779-1799}
\email{thkim@ntu.edu.sg}
\affiliation{\institution{Nanyang Technological University}\city{Singapore}\country{Singapore}}

\author{Yongjun Park}
\orcid{0000-0003-3725-0380}
\email{yongjunpark@yonsei.ac.kr}
\affiliation{\institution{Yonsei University}\city{Seoul}\country{South Korea}}

\author{Seongmin Hong}
\orcid{0009-0005-7940-8221}
\email{sm.hong@hyperaccel.ai}
\affiliation{\institution{HyperAccel}\city{Seoul}\country{South Korea}}

\author{Joo-Young Kim}
\authornote{Corresponding author.}
\orcid{0000-0003-1099-1496}
\email{jy.kim@hyperaccel.ai}
\affiliation{\institution{HyperAccel}\city{Seoul}\country{South Korea}}

\renewcommand{\shortauthors}{Yang et al.}

\input{0_Text/0_abstract}




\maketitle

\input{0_Text/1_introduction}
\input{0_Text/2_background}
\input{0_Text/3_motivation}
\input{0_Text/4_architecture}
\input{0_Text/5_top_architecture}
\input{0_Text/6_implementation}
\input{0_Text/7_evaluation}
\input{0_Text/8_related_work}
\input{0_Text/9_conclusion}

\input{main.bbl}

\end{document}

%% file: 0_Text/0_abstract.tex
\begin{abstract}

The number formats that accelerators execute natively are fixed at fabrication, yet large language model (LLM) checkpoints are released with MXFP4 weights in the Microscaling (MX) formats standardized in 2023.
FPGAs adopt new formats through bitstreams.
However, to the best of our knowledge, no deployed FPGA LLM accelerator executes MXFP4 weights alongside BF16 products.
We characterize the decoding of MXFP4 checkpoints, whose layers alternate between MXFP4$\times$BF16 products and BF16$\times$BF16 products of attention and BF16 layers.
With a BF16 KV cache, BF16$\times$BF16 products take 56\%-77\% of the datapath time, during which a separate MXFP4 unit idles.
We present \sysname, whose unified dot-product unit (UDPU) executes both product types on one DSP array and reduction pipeline.
First, per-group pre-alignment keeps this pipeline at the BF16 width and changes perplexity by at most 0.7\%.
Second, pre-adder packing places two signed MXFP4$\times$BF16 products in one DSP48E2 multiplier with an exact correction.
On an Alveo U55C, the UDPU needs 23\% fewer DSP slices and 36\% fewer LUTs than separate MXFP4 and BF16 units of equal throughput.
On this board, MXFP4 weights speed up decoding by 1.15$\times$-1.87$\times$ over BF16 weights, and an NVIDIA A40 GPU decodes faster but spends 0.92$\times$-2.01$\times$ the board's energy per token.

\end{abstract}

%% file: 0_Text/1_introduction.tex
\section{Introduction}
\label{sec-intro}

The number formats that an accelerator executes natively are fixed when it is fabricated, yet those of large language model (LLM) inference~\cite{grattafiori2024llama,openai2025gptoss120bgptoss20bmodel} now change within the service life of deployed hardware.
LLM inference has moved from the FP16 of the 2020 A100 GPU to FP8 in 2022 and to the 4-bit Microscaling (MX) formats standardized in 2023~\cite{9361255,micikevicius2022fp8,ocp_mx_spec_v1}.
MXFP4 has become a production format, because open-weight LLMs from the 20B-parameter GPT-OSS~\cite{openai2025gptoss120bgptoss20bmodel} to the 2.8T-parameter Kimi K3~\cite{moonshot2026kimik3} release most of their weights in MXFP4.
Narrower formats translate directly into decode throughput, because small-batch LLM inference streams the entire weight set from memory for every generated token, which leaves memory bandwidth as the binding constraint~\cite{kim2023squeezellm}.
This bottleneck has been addressed from both sides, as weight quantization reduces the bytes per weight~\cite{frantar2022gptq,lin2024awq} and inference accelerators from GPUs to proprietary LLM chips raise the bandwidth behind the datapath~\cite{9361255,park2025reviewproprietaryacceleratorslarge}.

However, a chip cannot natively execute a format that appears after its fabrication.
A model released in a newer format is upcast to a supported format such as BF16 when it is loaded~\cite{openai2025gptoss120bgptoss20bmodel}, which gives up the 4-bit weight bytes that motivated the format.
Alternatively, software kernels can keep the bytes by expanding the weights inside every kernel~\cite{hf2026mxfp4}, but every kernel then spends instructions converting the weights to BF16 on a datapath that was never built for MXFP4.
FPGA accelerators for LLM inference~\cite{hong2022dfx,moon2024lpu,zeng2024flightllm,chen2024spatial,ma2026cdllm} build custom datapaths next to high-bandwidth memory (HBM) and can adopt a new format after deployment by loading a new bitstream.
Yet none of their deployed datapaths executes MXFP4 alongside the BF16 work the model still requires.

LLM activations stay in FP16 or BF16 even when the weights are quantized, because they span a dynamic range that 4-bit or per-tensor integer formats cannot cover without accuracy loss~\cite{zhang2024integer}.
Therefore, weight quantization produces an asymmetric product in which a narrow weight meets a BF16 activation.
Floating-point (FP) accelerators built for quantized weights execute this product by dequantizing the weight in front of a BF16 multiplier~\cite{9361255}.
Even recent GPUs that execute MX formats natively~\cite{nvidia_blackwell_datasheet,mishra2025recipespretrainingllmsmxfp8} dequantize this product first, because their tensor cores require the same low-precision format on both operands~\cite{nvidia2026ptxisa}.

\input{1_Figures/TRETS/introduction/fig-llm-inference-float}

In contrast, on an FPGA, a bitstream can provide a datapath that multiplies the MXFP4 weight by the BF16 activation directly.
As Figure~\ref{fig-llm} shows, linear layers with MXFP4 weights produce MXFP4$\times$BF16 products, which define the \emph{MX-FP mode}.
Attention and layers such as the LM head that quantized checkpoints keep in BF16~\cite{openai2025gptoss120bgptoss20bmodel} produce BF16$\times$BF16 products, which define the \emph{FP-FP mode}.
Supporting MXFP4 decides which products the fabric can execute but not how the two modes share it, because they alternate within every layer of a token.
Across three open-weight LLMs, the FP-FP mode takes 56\%-77\% of the time the datapath spends on a decode token, as Section~\ref{sec:mode-share} shows.

Despite this proportion, the FP-INT units proposed for LLMs~\cite{jang2024figna,park2025figlut,mo2025lut} leave the BF16$\times$BF16 work to a separate unit, so each of the two units idles while the other mode runs.
Sizing the two units to the mode ratio helps only when both run at once, which the sequential phases of single-batch decoding preclude.
In contrast, a single datapath for both modes stays busy in every phase.
Such datapaths exist as ASICs with FP8 or block floating-point activations~\cite{wang2026harmonia,zhou2026mxglut} and as FPGA units~\cite{yu2026xtramac,he2026skipopu}, but neither FPGA unit exploits the block-shared scale of MX during alignment.
Aligning MX groups and BF16 products in one datapath on an FPGA fabric raises two challenges.

First, an alignment datapath that aligns every MX-FP product individually needs four times its FP-FP logic, because an MXFP4 vector holds four times as many elements as a BF16 vector of the same width.
On an FPGA, exponent comparison and alignment shifting are built from LUTs.
Because the other parts of an accelerator such as its vector unit and controllers also use LUTs, the LUT cost of a dot-product unit limits how many units fit just as its DSP cost does.
An MX-FP operand pair carries one shared scale per block and narrow elements, whereas an FP-FP pair carries an exponent per element.
Aligning a group of products at a time keeps the datapath at the FP-FP width, but the products of a group share only the weight scale, because their BF16 activations carry different exponents.

Second, an MXFP4$\times$BF16 product uses only a small part of a fixed-width DSP slice~\cite{xilinx2021ug579}, so one product per slice leaves most of each slice idle in the MX-FP mode.
In a single datapath, the same slices must hold several MXFP4$\times$BF16 products in the MX-FP mode and serve BF16$\times$BF16 products in the FP-FP mode.
Packing narrow integer products into one DSP is a standard technique~\cite{fu2017wp486,chen2024spatial}, but a BF16 weight mantissa needs a wider field than an MXFP4 weight.
Therefore, one DSP array must either switch its field layout between modes or divide each BF16 mantissa into fields of MXFP4 width.

To tackle these challenges, we present \sysname, whose \emph{unified dot-product unit} (UDPU) executes both modes on one DSP array and one alignment-and-summation datapath that we call the \emph{reduction pipeline}.
The key insight is that the two modes differ only in operand structure, because both compute a mantissa product per element pair followed by an aligned sum.
Therefore, each mode keeps its own \emph{product stage} that prepares the operands and forms the products on the unified DSP array, while the reduction pipeline is common to both modes.

\textbf{Unified alignment-and-summation datapath.}
Per-group pre-alignment lets the two modes share one reduction pipeline at the FP-FP width.
Pre-summation in the MX-FP product stage collapses each group of four products into one sum and one exponent, so a group enters the pipeline in place of one BF16 product.
Thus, the MX-FP mode multiplies four times as many elements per cycle as the FP-FP mode without widening the pipeline.

\textbf{Unified DSP array through pre-adder packing and bit slicing.}
One DSP array executes both modes, because the same packing serves MXFP4$\times$BF16 products and bit-sliced BF16$\times$BF16 products.
The DSP48E2 pre-adder packs two signed MXFP4$\times$BF16 products into one multiplier without a dequantizer, and a single correction term keeps both products exact.
BF16$\times$BF16 products are bit-sliced onto the same DSP slices at one product per two multipliers, a quarter of the MX-FP product rate.
Because BF16 holds about four times the bits of MXFP4, one array almost fully utilizes the memory bandwidth in both modes.

\textbf{Evaluation from DSP slices to end-to-end LLM inference.}
Under the platform's sequential execution, the UDPU matches the per-mode throughput of separate MXFP4 and BF16 units with 23\% fewer DSP slices and 36\% fewer LUTs.
This saving does not depend on how long the FP-FP mode runs.
We implement a dequantizing BF16 datapath and three split datapaths on the same fabric at the same per-mode throughput and with at least the accuracy of the UDPU.
On one deployed U55C board, MXFP4 weights speed up end-to-end decoding by 1.15$\times$-1.87$\times$ over BF16 weights on the same bitstream.
An NVIDIA A40 GPU decodes faster but spends 0.92$\times$-2.01$\times$ the board's energy per token.
Pre-alignment changes perplexity by at most 0.7\%, which stays below the cost of the 4-bit weights themselves on every dense model.

Overall, the UDPU adds the FP-FP mode to an MX datapath at no DSP cost and 17.5\% more LUTs.
The UDPU enables a deployed board to execute a new format natively, whereas fixed silicon of the same generation supports the format only through software dequantization.

To summarize, our key contributions are as follows.

\begin{itemize}[labelindent=0.0em,nolistsep,leftmargin=1.3em]
\item We characterize the decoding of MXFP4 checkpoints and show that BF16$\times$BF16 products take 56\%-77\% of the datapath time, so a separate MXFP4 unit idles for most of each token.
\item We present the UDPU, whose per-group pre-alignment lets two mode-specific product stages share one reduction pipeline at the FP-FP width, and we bound the truncation error of this pre-alignment.
\item We map both modes onto one DSP48E2 array, packing two signed MXFP4$\times$BF16 products per DSP with an exact correction and bit-slicing BF16$\times$BF16 products onto the same slices.
\item We compare the UDPU with four alternative datapaths on the same fabric, and we measure the MX-FP gain on the deployed bitstream and the accuracy with an emulator whose alignment is bit-exact with the RTL.
\end{itemize}

%% file: 1_Figures/TRETS/introduction/fig-llm-inference-float.tex
\begin{figure}[t]
\centering
\includegraphics[width=1.0\linewidth]{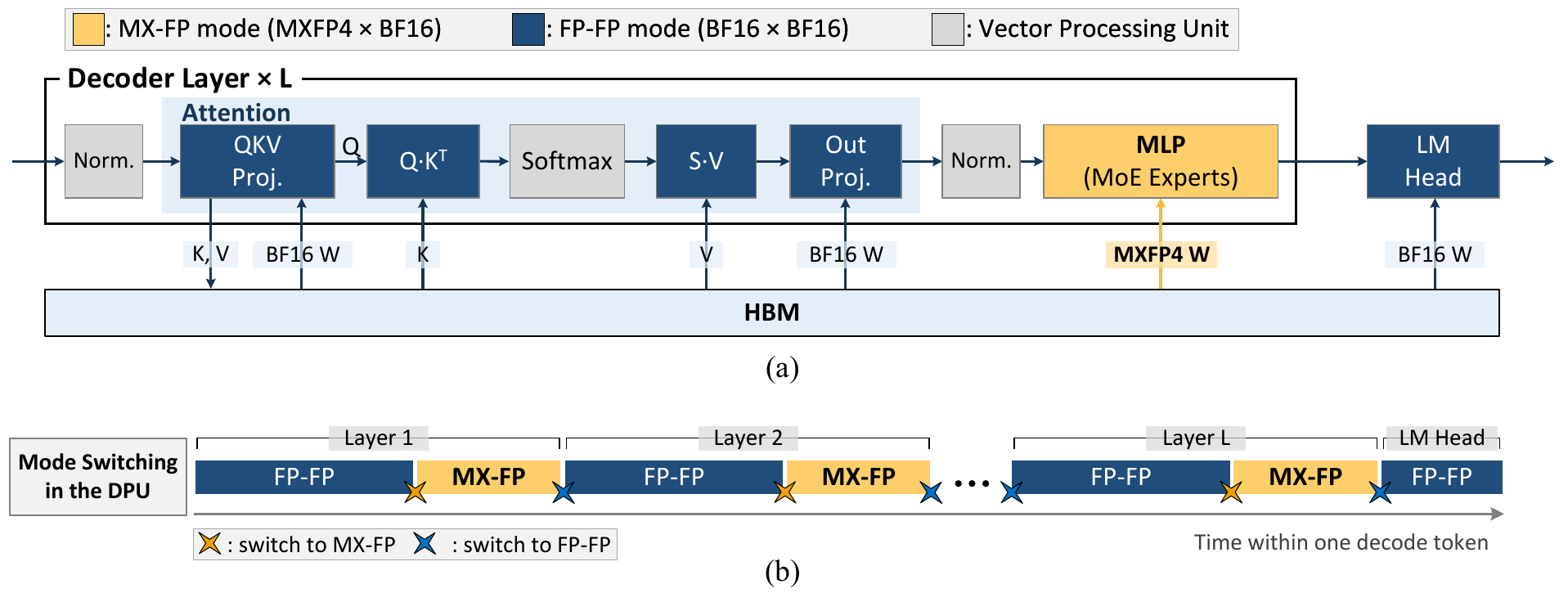}
\Description{Block diagram of a decoder layer whose products are colored by execution mode, and a timeline of one layer in which the dot-product units switch between the two modes twice.}
\caption{Execution modes in one decode token. (a) A decoder layer with each product colored by its mode. (b) Data dependence orders the two modes, so the dot-product units switch modes twice per layer.}
\label{fig-llm}
\end{figure}

%% file: 0_Text/2_background.tex
\section{Background}
\label{sec-back}

\input{1_Figures/TRETS/background/fig-mx-float}

\subsection{Microscaling Data Format}
\label{sec-back_mx}

The Open Compute Project (OCP) MX specification defines block floating-point formats in which 32 elements share one 8-bit scale~\cite{ocp_mx_spec_v1,rouhani2023microscalingdataformatsdeep}.
As Figure~\ref{fig-mx}(b) shows, the scale uses the E8M0 encoding and is thus a power of two without a mantissa, whereas each element keeps its own narrow encoding.
MXFP4 is the narrowest member of the family and encodes an element as E2M1, which holds a sign bit, two exponent bits and one mantissa bit~\cite{ocp_mx_spec_v1}.
Every E2M1 magnitude is a multiple of one half and at most 6, so doubling an element turns it into a signed 5-bit integer.
BF16 is the activation format of both modes, and Figure~\ref{fig-mx}(a) shows its sign bit, 8-bit exponent and 7-bit mantissa with an implicit leading one.
Its exponent is as wide as an E8M0 scale.
Therefore, an MXFP4 element costs 4.25 bits including its 1/32 part of the scale, while a BF16 value costs 16 bits.
Four MXFP4 elements occupy the width of one BF16 value, which Section~\ref{sec-arch} calls the widening factor of the MX-FP mode.
A weight tensor stored in MXFP4 shrinks to a little over one quarter of its BF16 footprint, a ratio on which the memory-traffic analysis of Section~\ref{sec-motiv} builds.

MX is a hardware-native form of group quantization~\cite{ocp_mx_spec_v1,darvish2023shared}, and Equation~\ref{eq:mx_quant} is the conversion the specification requires as a minimum for a block of values $v_i$.
\begin{equation}
\label{eq:mx_quant}
X = 2^{\lfloor \log_2 \max_i |v_i| \rfloor - e_{\max}}, \qquad p_i = \mathrm{Q}\!\left(v_i / X\right)
\end{equation}
The shared exponent is the exponent of the largest magnitude in the block less $e_{\max}$, the exponent of the largest power of two the element format represents, which is 2 for E2M1.
$\mathrm{Q}$ rounds the scaled value to the nearest element with ties to even and saturates at the largest element, and the specification permits other conversion algorithms~\cite{ocp_mx_spec_v1}.
Integer group quantization pairs an arbitrary FP scale with INT4 elements~\cite{frantar2022gptq,lin2024awq}, whereas the MX scale is a power of two.
Therefore, applying the MX scale takes an exponent addition or a shift instead of a multiplication.
Because the scale is common to a block, a block dot product factors into a sum of element products followed by one scale application~\cite{ocp_mx_spec_v1}, which lets Section~\ref{sec-arch} multiply MXFP4 elements before their scale is applied.

\subsection{Floating-Point and FP-INT Dot-Product Units}
\label{sec-back_dpu}

\input{1_Figures/TRETS/background/fig-dpu-float}

A dot product multiplies two length-$n$ vectors element by element and accumulates the products, as in Equation~\ref{eq:dot_vector}.
\begin{equation}
\label{eq:dot_vector}
\mathrm{DotProduct}(\mathbf{a}, \mathbf{b}) = \sum_{i=0}^{n-1} a_i b_i
\end{equation}
As Figure~\ref{fig-dpu}(a) shows, an FP dot-product unit (DPU) executes it in three stages called multiplication, alignment and summation~\cite{abdelaziz2021rethinking,4751896,6545890}.
The multiplication stage computes $k$ products per cycle, each as a mantissa product and an exponent sum, where $k$ is the width of the unit.
The alignment stage finds the largest of the $k$ exponents and right-shifts every mantissa by its distance from that exponent, so all products share one exponent.
The summation stage adds the aligned mantissas in an adder tree.
A vector longer than $k$ passes through the unit over several cycles, so an accumulator holds the partial sum and its exponent and the result is normalized at the end.
A DPU with $k = 1$ is a conventional FP multiply-accumulate unit~\cite{5661880}, and every stage grows with $k$.
The alignment and summation stages are built from LUTs on an FPGA, whereas the multipliers map to the DSP slices that Section~\ref{sec-back_dsp} describes.

Figure~\ref{fig-dpu}(b) shows three methods for a product between a low-precision weight and an FP activation, which rely on a dequantizer, an integer unit or a lookup table.
The first method dequantizes each MXFP4 weight to the BF16 activation format in front of an unmodified FP DPU, so every weight the unit consumes passes through a dequantizer.
The second method, taken by FIGNA, first aligns the mantissas of an activation vector to its largest exponent~\cite{jang2024figna}.
An integer unit then multiplies them by the weights and sums the products without per-product alignment.
It needs no dequantizer, and a bounded truncation of the aligned mantissas keeps its accuracy at that of FP arithmetic.
The third method, taken by FIGLUT and LUT Tensor Core, replaces the multiplier with a table of partial products that is built from the activations and indexed by the bits of the weight~\cite{park2025figlut,mo2025lut}.
FIGLUT also offers an integer variant that pre-aligns the activations in the same way~\cite{park2025figlut}.
These three methods were proposed for integer weights, a setting that FIGNA and FIGLUT call the FP-INT product~\cite{jang2024figna,park2025figlut}.
We extend that name to MXFP4, whose E2M1 elements enter the datapath as signed 5-bit integers once a power of two is moved into the scale as Section~\ref{sec-back_mx} describes.
The first method keeps one FP DPU for both modes, whereas the second and third methods execute only the FP-INT product and need a separate BF16 DPU for the FP-FP mode.

\input{1_Figures/TRETS/background/fig-packing-float}

\subsection{DSP48E2 and DSP Packing}
\label{sec-back_dsp}
The DSP48E2 is the hardened arithmetic slice of the AMD UltraScale architecture~\cite{xilinx2021ug579}.
As Figure~\ref{fig-packing}(a) shows, each slice evaluates Equation~\ref{eq:dsp} over its input ports.
\begin{equation}
\label{eq:dsp}
P = (A + D) \times B + C
\end{equation}
The 27-bit input $D$ and the lower 27 bits of input $A$ feed the pre-adder, and $B$ is the 18-bit input of the $27 \times 18$-bit two's-complement multiplier.
Input $C$ drives the 48-bit post-adder, which also serves as an accumulator.
Because the DSP48E2 is the only hard multiplier on the device, a design that multiplies on DSP slices computes at most the slice count times the products per slice in each cycle.

DSP packing raises the products per slice by executing two narrow multiplications within a single $27 \times 18$-bit multiplier~\cite{fu2017wp486,chen2024spatial}.
Two weights that share an activation enter ports $A$ and $D$, and one of them is shifted left by $s$ bits.
The pre-adder then combines them into one wide operand, so the shared activation in port $B$ multiplies both weights at once as Figure~\ref{fig-packing}(b) shows.
Consequently, the 45-bit product holds the two products in disjoint bit fields if $s$ is at least the width of the lower product.
INT8 packing places two $8 \times 8$ multiplications in one DSP slice with $s = 18$~\cite{fu2017wp486}.
The scheme also extends to narrower and asymmetric precisions such as 4-bit weights with 8-bit activations~\cite{chen2024spatial}.
With signed operands, a negative lower product borrows one from the upper field, so adding $2^s$ to the result restores the upper product.

%% file: 1_Figures/TRETS/background/fig-mx-float.tex
\begin{figure}[!t]
\centering
\includegraphics[width=\linewidth]{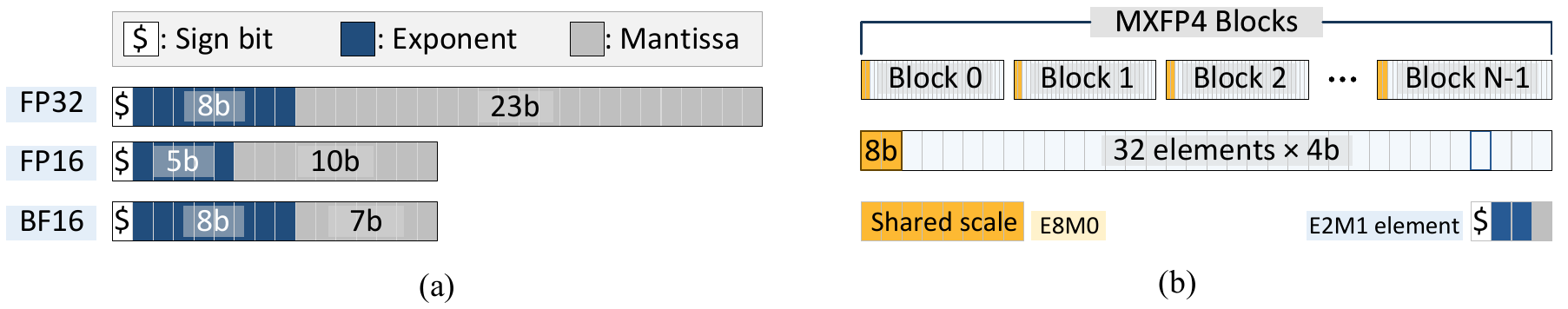}
\Description{Bit layouts of FP32, FP16, and BF16, and an MXFP4 block of 32 four-bit elements that share one 8-bit scale.}
\caption{(a) FP32, FP16, and BF16 bit layouts. (b) MXFP4 blocks of 32 E2M1 elements with one shared E8M0 scale.}
\label{fig-mx}
\end{figure}

%% file: 1_Figures/TRETS/background/fig-dpu-float.tex
\begin{figure}[!t]
\centering
\includegraphics[width=\linewidth]{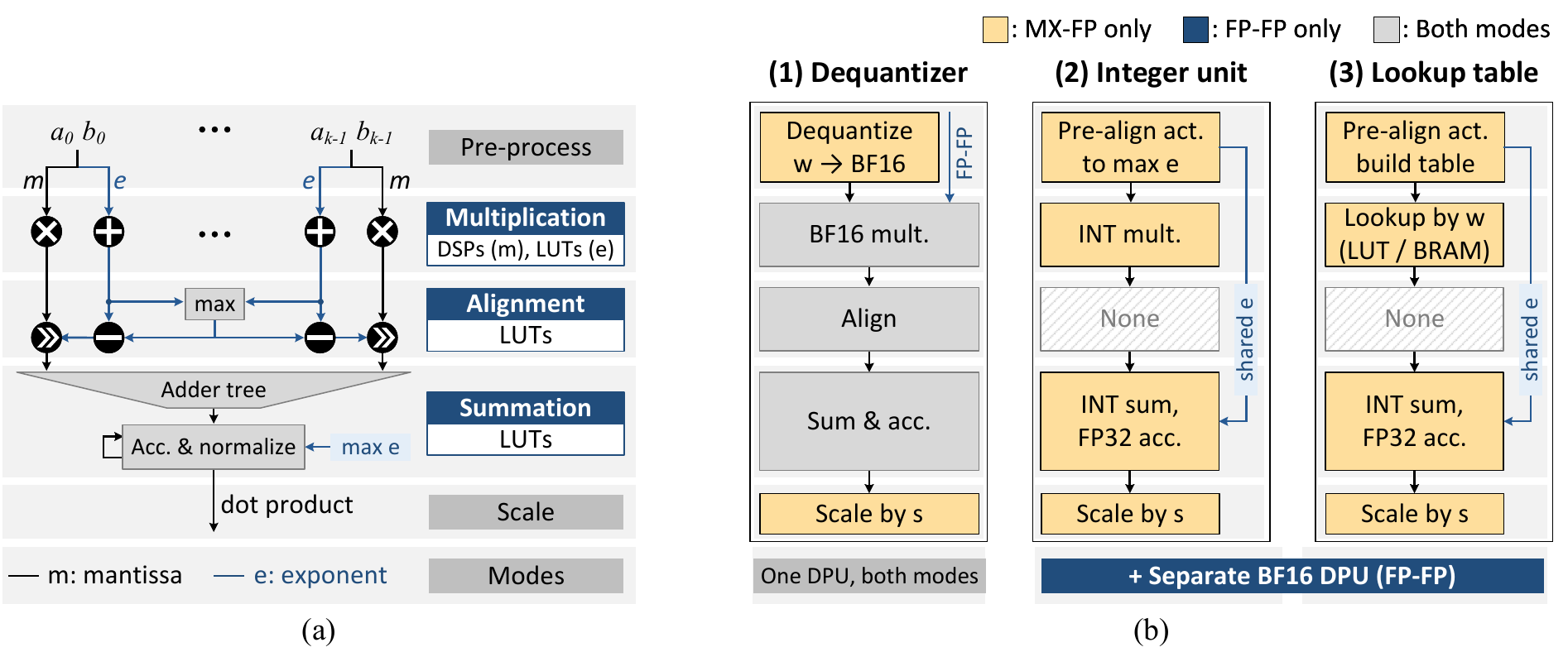}
\Description{Block diagram of a floating-point dot-product unit with multipliers, alignment shifters, and an adder tree, next to three datapaths for a product of an MXFP4 weight and a BF16 activation.}
\caption{(a) A $k$-wide FP dot-product unit, with mantissa products on DSP slices and alignment and summation on LUTs. (b) Three methods for an MXFP4$\times$BF16 product, colored by the mode each block serves.}
\label{fig-dpu}
\end{figure}

%% file: 1_Figures/TRETS/background/fig-packing-float.tex
\begin{figure}[!t]
\centering
\includegraphics[width=\linewidth]{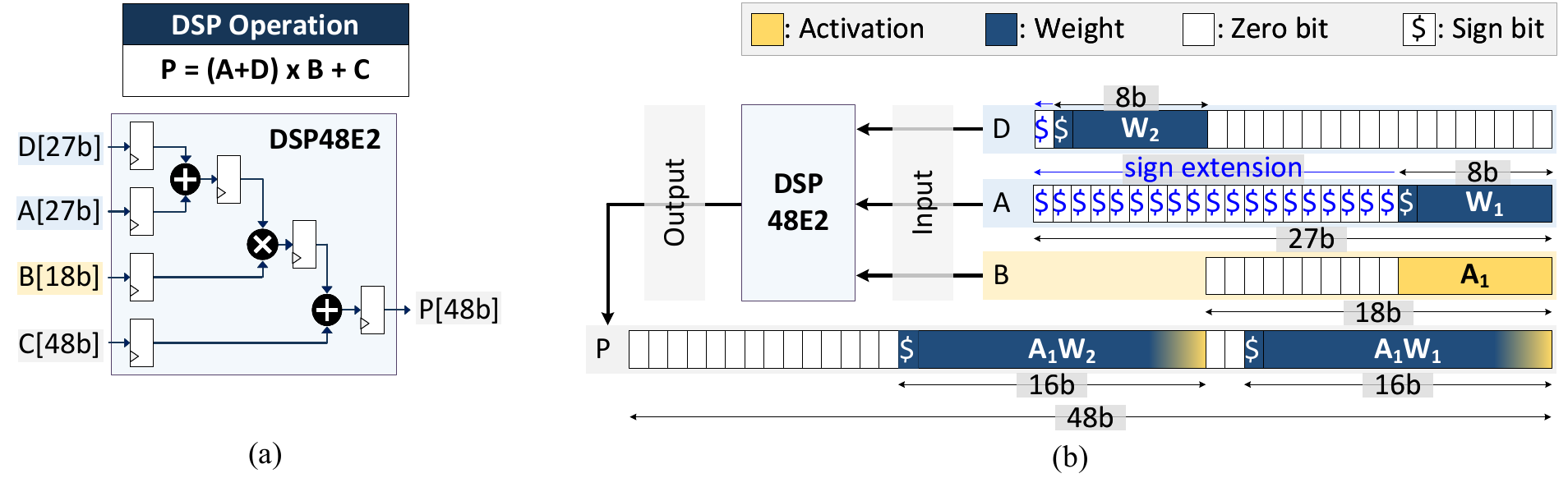}
\Description{Diagram of a DSP48E2 slice with its pre-adder, multiplier, and post-adder, and the bit fields of two packed 8-bit products in its output.}
\caption{DSP packing on a DSP48E2 slice, adapted from~\cite{yang2024acane}. (a) The slice evaluates Equation~\ref{eq:dsp}. (b) Two 8-bit weights that share an activation enter $A$ and $D$, so $P$ holds both products in disjoint fields.}
\label{fig-packing}
\end{figure}

%% file: 0_Text/3_motivation.tex
\section{Characterization and Motivation}
\label{sec-motiv}

\input{1_Figures/TRETS/motivation/tab-kv-precision}

\subsection{Precision Asymmetry between Weights and Activations}
\label{sec:precision}

The two operands of an LLM product tolerate 4-bit quantization very differently, so MXFP4 suits the weights but not the activation-activation products.
Weights are static tensors whose distribution is known before deployment, so post-training quantization chooses their scales offline and can reshape them against calibration data~\cite{frantar2022gptq,lin2024awq}.
In contrast, activations are produced at run time with outlier channels and token-dependent ranges~\cite{dettmers2022gpt3,bondarenko2021understandingovercomingchallengesefficient,xiao2023smoothquant}.
In attention, both operands are activations, because a query multiplies the cached keys and the attention probabilities multiply the cached values.
MXFP4 checkpoints also keep some layers in BF16, such as the attention projections and LM head of GPT-OSS-20B~\cite{openai2025gptoss120bgptoss20bmodel}, so their products stay BF16.
Therefore, a checkpoint with MXFP4 weights still multiplies BF16 by BF16 in attention and in these layers.

Table~\ref{tab-kv-precision} shows this asymmetry with the RULER long-context benchmark~\cite{hsieh2024ruler}.
MXFP4 MLP weights with BF16 attention stay within 1.1 points of the BF16 model from 1K to 16K on Llama-3.1-8B and Qwen3-8B.
GPT-OSS-20B has no BF16 release, so its rows start from MXFP4 weights.
Attention with its queries, keys, and values in MXFP4 loses 15 to 65 points against BF16 attention at 16K and 35 to 71 points at 32K on the three models.
This loss overstates that of a quantized KV cache, because the queries are quantized as well.
Therefore, a model keeps its accuracy with MXFP4 weights but loses a large part of it with MXFP4 attention.

Prior work keeps a 4-bit KV cache accurate, but its methods need more than the MX-FP mode provides.
These methods rely on asymmetric integer groups or on tokens kept in a wider format~\cite{liu2024kivi,hooper2024kvquant,lin2025qserve,kim2025oaken,su2025kvsink,chen2024prefixquant}.
MXFP4 offers neither the zero points of these integer formats nor a wider format for selected tokens.
An MXFP4 KV cache must also quantize each key and value as its token arrives, and its value blocks must lie along the token axis that the probabilities sum over.
Therefore, this paper keeps the KV cache in BF16, and a 4-bit KV cache stays outside its scope.

Attention in an 8-bit format keeps the accuracy, but it does not remove the FP-FP mode.
FP8 and MXFP8 attention lose at most 1.2 points against BF16 attention with the same MXFP4 weights.
Their keys and values widen to BF16 without loss, so attention on them still runs in the FP-FP mode without new multipliers.
Only a mode that multiplies the 8-bit keys and values without widening them takes attention out of the FP-FP mode, but it needs dedicated hardware.
That is, each FP8 or MXFP8 element carries its own exponent and does not fit the 5-bit weight field of the MX-FP mode, so every product must be aligned separately.
This hardware is added to the FP-FP mode and does not replace it, because the BF16 layers still multiply BF16 by BF16.
Therefore, an MX datapath requires the FP-FP mode regardless of the precision that attention adopts.

\textbf{Takeaway.} \textit{An MX datapath must also execute the FP-FP mode, because MXFP4 checkpoints keep BF16 layers and attention loses its accuracy in MXFP4.}

\input{1_Figures/TRETS/motivation/fig-mode-share-float}

\subsection{Interleaving Two Execution Modes within a Token}
\label{sec:mode-share}

The two modes alternate within every decoder layer at a granularity that FPGA reconfiguration cannot follow, so one bitstream must execute both modes.
Each linear layer executes in the mode of its weights, and both attention products read the BF16 KV cache in the FP-FP mode.
We give each dense model the layout of GPT-OSS-20B, with its MLP weights in MXFP4 and its attention projections in BF16.
In the simulated Llama-3.1-8B layer of Figure~\ref{fig-mode-share}(a), the DPUs execute the QKV projection, the attention and the output projection in the FP-FP mode before the MLP in the MX-FP mode.
This order follows from data dependence rather than from our schedule, so the two modes never execute concurrently and each decoder layer divides into single-mode phases.
Each phase lasts at most tens of thousands of cycles, far shorter than the time to load a bitstream.
Consequently, the datapath of one bitstream must switch between the two modes twice in every decoder layer.

An MXFP4 model reads more than half of its decode bytes in the FP-FP mode, although MXFP4 holds most of its weights.
Each decode token reads every weight once and the KV cache of all earlier positions, at 16 bits per BF16 element and 4.25 bits per MXFP4 weight.
For three open-weight models~\cite{grattafiori2024llama,yang2025qwen3,openai2025gptoss120bgptoss20bmodel}, the FP-FP mode reads 56\%-70\% of these bytes at every position up to 16K, because a BF16 element occupies 3.76$\times$ the bits of an MXFP4 element.
Batching only raises this proportion, because it shares the weight reads across requests but not the KV reads.
As a result, a datapath for MXFP4 models must run the FP-FP mode at the full bandwidth as well, because this mode streams most of the decode bytes.

The FP-FP mode takes most of the DPU time of a token as well as most of its bytes.
In the RTL simulation of Section~\ref{sec-eval-setup}, this time proportion is 56\%-77\% over the three models up to 8K and reaches 83\% at 16K, as Figure~\ref{fig-mode-share}(b) shows.
At the first token position, the time proportion of each model matches its byte proportion, because both modes stream near the peak HBM bandwidth.
This match means that the proportion at the first position holds for any datapath that saturates the bandwidth in both modes, not only for the one we simulate.
At later positions, the time proportion rises above the byte proportion, because only the attention lengthens and the accelerator rereads each KV element for every query head.
Therefore, the FP-FP mode takes more than half of the DPU time at every position, with a proportion that shifts with the model and the position.

\textbf{Takeaway.} \textit{One datapath must run both modes at the full bandwidth, because they alternate in every decoder layer and the FP-FP mode takes most of the DPU time.}

\subsection{Limits of Dequantizing and FP-INT Datapaths}
\label{sec:split-unified}

Three datapaths can provide both modes, namely an FP-only extension, a split datapath, and a unified datapath.
They differ in how they spend DSP slices and LUTs.
The FP-only extension, the approach of FP accelerators for quantized weights~\cite{9361255}, dequantizes the MXFP4 weights in front of an unmodified BF16 DPU.
The split datapath pairs an MX-FP lane built from an FP-INT unit~\cite{jang2024figna,park2025figlut} with a separate BF16 lane.
The unified datapath instead executes both modes on one DSP array and one alignment and summation path.
Because each operand that a decode token reads enters one product, the HBM bandwidth sets the product rate of each mode.
Therefore, the FP-FP mode needs only about a quarter of the MX-FP product rate on any platform, because a BF16 operand holds 3.76$\times$ the bits of an MXFP4 weight.
We rank the options by DSP slices first, because the DSP48E2 is the sole hard multiplier.
A unified array sized for two packed MX-FP products per DSP leaves two slices for every BF16 product, so the FP-FP mode needs no additional DSP slices.
Therefore, the unified datapath sets the DSP count that the other two must match.

Under the sequential schedule of Section~\ref{sec:mode-share}, a split with DSP multipliers loses to the unified datapath in DSP slices or in token time, because one of its lanes idles in every phase.
At 2K, its MX-FP lane idles for 62\%-69\% of the DPU time and its BF16 lane for 31\%-38\%.
A split with both lanes at the full bandwidth needs 1.25$\times$ the multiplier DSP slices, and halving its BF16 lane slows a token by 30\%-62\%.
Partial reconfiguration cannot swap the lanes, because loading even the smaller lane takes over 10~ms~\cite{amd2024ug909} whereas a phase lasts less than 0.3~ms.
A split sized to the largest proportion of each mode in Figure~\ref{fig-mode-share}(b) saves DSP slices only under an overlapped schedule.
Consequently, among the datapaths that multiply in DSP slices, only the FP-only extension remains a rival to the unified datapath.

An FP-only extension with packed BF16 units ties the unified datapath in multiplier DSP slices but spends most of its LUTs on aligning and summing products at the BF16 width.
An MXFP4 vector holds four times the elements of a BF16 vector of equal width, so aligning every MX-FP product takes four times the shifters of the FP-FP mode.
Therefore, the first requirement for saving LUTs is a shared alignment path that stays at the FP-FP width.
The second requirement is one DSP packing for both modes, although a BF16 mantissa needs a wider field than an MXFP4 element.
Consequently, a unified datapath improves on the FP-only extension only if it meets both requirements.

\textbf{Takeaway.} \textit{Split FP-INT datapaths idle DSP slices and dequantizing datapaths spend LUTs on BF16-width alignment, so a unified datapath needs an FP-FP-width alignment path and one DSP packing for both modes.}

%% file: 1_Figures/TRETS/motivation/tab-kv-precision.tex
\begin{table}[t]
\centering
\caption{RULER accuracy (\%) over 12 tasks by the precision of the weights and of attention, with the queries, keys, and values of attention in one format. Each task has 100 samples up to 16K and 50 at 32K.}
\label{tab-kv-precision}
\small
\begin{tabular*}{\linewidth}{@{\extracolsep{\fill}}lllrrrrrr@{}}
\toprule
Model & Weights & Attention & 1K & 2K & 4K & 8K & 16K & 32K \\
\midrule
Llama-3.1-8B & BF16 & BF16 & 99.0 & 97.0 & 98.0 & 96.9 & 96.4 & 90.2 \\
 & \textbf{MXFP4} & \textbf{BF16} & 99.0 & 96.4 & 98.1 & 96.6 & 95.3 & 92.7 \\
 & MXFP4 & FP8 & 98.8 & 96.6 & 98.0 & 96.2 & 95.3 & 91.8 \\
 & MXFP4 & MXFP8 & 99.0 & 96.8 & 98.0 & 96.5 & 95.5 & 92.7 \\
 & MXFP4 & MXFP4 & 92.2 & 87.4 & 86.1 & 77.6 & 64.3 & 58.2 \\
\midrule
Qwen3-8B & BF16 & BF16 & 98.5 & 97.6 & 97.8 & 95.8 & 95.4 & 93.5 \\
 & \textbf{MXFP4} & \textbf{BF16} & 98.2 & 97.2 & 96.9 & 95.3 & 95.7 & 92.6 \\
 & MXFP4 & FP8 & 98.3 & 97.4 & 96.8 & 95.4 & 95.7 & 93.3 \\
 & MXFP4 & MXFP8 & 98.4 & 97.4 & 97.2 & 95.9 & 95.5 & 91.9 \\
 & MXFP4 & MXFP4 & 97.0 & 95.4 & 95.0 & 89.4 & 80.3 & 47.6 \\
\midrule
GPT-OSS-20B & \textbf{MXFP4} & \textbf{BF16} & 98.8 & 95.6 & 97.8 & 94.4 & 93.4 & 90.6 \\
 & MXFP4 & FP8 & 99.1 & 95.3 & 97.5 & 94.5 & 93.3 & 89.9 \\
 & MXFP4 & MXFP8 & 99.0 & 95.5 & 97.5 & 94.2 & 93.3 & 89.4 \\
 & MXFP4 & MXFP4 & 54.9 & 41.4 & 42.4 & 39.8 & 28.3 & 19.8 \\
\bottomrule
\end{tabular*}
\end{table}

%% file: 1_Figures/TRETS/motivation/fig-mode-share-float.tex
\begin{figure}[t]
\centering
\includegraphics[width=\linewidth]{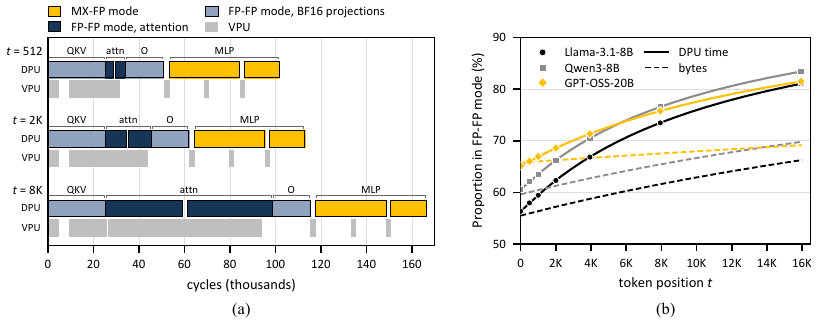}
\Description{Stacked bars of the decode time of one layer by mode at three token positions, and line plots of the proportion of time and bytes in the FP-FP mode against the token position for three models.}
\caption{Decode time of the two modes in RTL simulation. (a) One Llama-3.1-8B layer with its MLP quantized to MXFP4 at three token positions. (b) Proportion of DPU time (solid) and bytes (dashed) in FP-FP mode.}
\label{fig-mode-share}
\end{figure}

%% file: 0_Text/4_architecture.tex
\section{\sysname Compute Engine}
\label{sec-arch}

\subsection{Overview of the \sysname Compute Engine}

\input{1_Figures/TRETS/architecture/fig-arch-udpu-float}

Building on the three takeaways of Section~\ref{sec-motiv}, \sysname follows one design principle, \emph{which is to share every stage that the two modes have in common from the DSP multipliers to the accumulator}.
Both modes compute a mantissa product per element pair followed by an aligned sum and differ only in operand structure, so this principle lets them share most of the datapath.
Therefore, \sysname builds its compute engine around the UDPU, which runs both modes on one reduction pipeline and places the product stages of both modes on one DSP array.
The DSP array denotes the DSP48E2 slices that compute the mantissa products of both modes.
The MX-FP product stage pre-sums every four products into a sum with one exponent, so the reduction pipeline stays at the FP-FP width in both modes, which addresses the first challenge of Section~\ref{sec-intro}.
The DSP array packs two MXFP4$\times$BF16 products into each slice and bit-slices BF16$\times$BF16 products onto the same slices, so one array meets the product rate of both modes, which addresses the second challenge.
Because the FP-FP mode needs only about a quarter of the MX-FP product rate, the UDPU computes half a BF16 product per DSP slice so that both modes share one reduction pipeline per output lane.
This sharing adds the logic that selects between the two operand formats, which Sections~\ref{sec-arch-compose} and~\ref{sec-arch-dsp} describe.

Figure~\ref{fig:arch-udpu} shows the organization of the compute engine, and the following subsections present the mechanisms behind these design decisions.
In the MX-FP mode, the pre-alignment stage of Section~\ref{sec-arch-prealign} first shifts each group of four BF16 activations to their largest exponent, so the group carries one exponent.
Each MXFP4 weight is decoded into a 5-bit signed fixed-point value, so its product with an activation does not wait for the shared scale, as Section~\ref{sec-arch-scale} explains.
The DSP array computes these products with the packing of Section~\ref{sec-arch-dsp}, and the product stage sums them per group and adds the shared scale to each group exponent.
The reduction pipeline then aligns the group sums by these exponents and accumulates them.
In the FP-FP mode, activations bypass pre-alignment, and Section~\ref{sec-arch-compose} shows how the product stage bit-slices each weight-side BF16 mantissa onto the same DSP array and feeds the products to the same reduction pipeline.
Section~\ref{sec-arch-numerics} bounds the output error of both modes, and Section~\ref{sec-platform} describes how \sysname delivers operands to the UDPU in these formats.

\subsection{Mode-Specific Product Stages on a Shared Reduction Pipeline}
\label{sec-arch-compose}

\input{1_Figures/TRETS/architecture/fig-udpu-detail-float}

As Figure~\ref{fig:udpu-detail} shows, the UDPU concentrates the differences between the two modes in its product stages, each of which delivers the same number of mantissas per lane and cycle to the reduction pipeline.
Each product stage comprises the operand path that feeds the UDPU, the group units that multiply and add the mantissas, and the exponent adders that add the operand exponents.
The two product stages share the group units and the exponent adders.
Each UDPU multiplies one activation vector by two weight rows at once, one per output lane, so every DSP slice serves both lanes.
Accordingly, each group unit comprises four such DSP slices and a dedicated two-level pre-summation tree for each lane, as the left of Figure~\ref{fig:udpu-detail} shows.
In the MX-FP mode, the operand path pre-aligns each group of four activations to a common group exponent and decodes each MXFP4 weight into a 5-bit signed fixed-point value.
For each lane, the group unit multiplies the four activation mantissas of a group by four weights from the same MXFP4 block, since the block size of 32 is a multiple of four.
Consequently, the four products share one group exponent and one scale, so the pre-summation tree adds them as integers into a 16-bit group sum without alignment.

In the FP-FP mode, the product stage executes BF16$\times$BF16 products on the same group units through bit slicing, so this mode adds no DSP slices to the UDPU.
It differs from the MX-FP product stage only in the operand path, the first adder level, and a bypass of the second level.
Since BF16 products carry individual exponents and thus cannot be pre-summed, each group unit computes only one BF16$\times$BF16 product per lane and cycle.
To fit the 5-bit weight port of the multipliers, the operand path splits each 9-bit signed weight-side BF16 mantissa into two 5-bit slices.
The upper slice contains the sign bit and the four most significant bits, whereas the lower slice contains the four least significant bits extended with a zero sign bit.
Shifting the upper slice left by four bits and adding the lower slice restores even a negative mantissa, because its four least significant bits always carry positive weight in two's complement.
Therefore, each BF16 product reaches the adders of the group unit as two partial products of the same activation mantissa.

After its four-bit shift, the upper partial product exceeds the width of the first-level adders, yet the group unit completes each BF16 product exactly without widening them.
To keep this width, the first adder computes only the lower 14 bits of the sum of both partial products and a carry.
The FP-FP mode repurposes the idle second adder of the first level to add the remaining upper bits and the carry one stage later, so the group unit accepts a BF16 product every cycle.
The joined 19-bit result represents the product exactly in its low 17 bits, since two mantissas of magnitude below 256 produce a product of magnitude below $2^{16}$.
Because each BF16 product occupies two of the four multipliers of a group unit and leaves the other two idle, the FP-FP mode runs at a quarter of the MX-FP product rate.

All of this mode-specific work serves one goal, to present the reduction pipeline with input of the same form in both modes.
Each group unit delivers one signed mantissa of at most 17 bits with its own exponent per lane and cycle in either mode, so the reduction pipeline serves both modes with the same hardware.
The pipeline aligns the mantissas to their largest exponent, sums them in an adder tree, and then normalizes and accumulates the result in floating point across cycles.
These steps reuse the modules of the BF16 DPU of Section~\ref{sec-back_dpu}.
The only mode-dependent element is a multiplexer that increases the largest exponent by a fixed offset before normalization in the MX-FP mode.
The offset corrects the position of the binary point, because both MX-FP operands carry fewer fraction bits than BF16 mantissas, as Section~\ref{sec-arch-scale} derives.

\subsection{Decoupling the Shared Scale from the Multipliers}
\label{sec-arch-scale}

The UDPU multiplies MXFP4 weights without dequantizing them to BF16, because neither of their two exponents enters the multipliers.
The operand path removes the element exponent by decoding each element into a 5-bit signed fixed-point value with one fraction bit.
The decoder shifts the two-bit significand left by up to two places, so each element becomes a multiple of one half between $-6$ and 6.
The shared scale remains the only exponent of the weights in a block, as the integer pre-summation of Section~\ref{sec-arch-compose} requires.
Because this scale is a power of two, the exponent adders apply it in parallel with the multipliers, whereas the dequantizer of Section~\ref{sec-back_dpu} applies it to every weight first.

In the MX-FP mode, the operand path delivers the shared scale to the exponent adders through the input of the BF16 weight exponent, so the scale requires no dedicated adders.
Because E8M0 uses the same bias of 127 as BF16, the subsequent normalization and accumulation handle the bias of every exponent sum identically in both modes.
The exponent adders also operate without a mode signal, since the operand path alone selects between the shared scale and the BF16 weight exponent.

A group sum still carries only 7 fraction bits, whereas the reduction pipeline reads each input as a BF16 product with 14.
Of these 7 bits, the pre-aligned activation of Section~\ref{sec-arch-prealign} keeps 6 and the decoded element keeps 1.
Therefore, the pipeline reads each group sum at $2^{-7}$ of its value.
The multiplexer of Section~\ref{sec-arch-compose} restores this factor by adding 7 to the largest exponent of each lane before normalization.
Adding this offset once per lane corrects every group sum of the lane, because all of them carry the same factor and the alignment shifts depend only on exponent differences.
Thus, apart from the pre-alignment truncation that Section~\ref{sec-arch-numerics} bounds, the MX-FP mode adds no error beyond that of the reduction pipeline.

\subsection{Per-Group Pre-Alignment}
\label{sec-arch-prealign}

The operand path aligns each group of four BF16 activations to one exponent, so the UDPU pre-sums their products as integers although each activation has its own exponent.
Each mantissa is truncated to its hidden bit and upper six fraction bits and then shifted right to the group exponent $\mathit{exp}_g$, the largest exponent of its group, as Algorithm~\ref{alg:prealign} lists.
With its sign applied, each aligned mantissa becomes an 8-bit MXINT8 element~\cite{ocp_mx_spec_v1}, whose format sets the six fraction bits.
Thus, the conversion follows Equation~\ref{eq:mx_quant} over a group of four with truncation in place of rounding.
Truncation keeps the pre-alignment logic small, because rounding to nearest even takes 40\% more LUTs.
Each element is then sign-extended to nine bits, so the group units receive it on the activation port of the FP-FP mode and pre-sum its products without alignment.

\input{1_Figures/TRETS/architecture/alg-prealign}

Pre-alignment converts each activation vector only once, before the shared activation buffer of Section~\ref{sec-platform-mx} stores it for reuse, so the conversion rate is independent of the product rate of the UDPUs.
Each buffer slot holds either one aligned group or one BF16 activation, which allows one buffer read to deliver four times as many activations in the MX-FP mode.
Because every lane multiplies the same activations by its own weights, all UDPUs that read the buffer share one aligned copy of the vector.
Therefore, pre-alignment adds only one conversion stage before the buffer, and the cost of this stage grows with neither the product rate nor the number of lanes.

Groups of four are the smallest that allow the MX-FP mode to share the reduction pipeline, and their small size limits the truncation of pre-alignment.
A lane computes four times as many products in the MX-FP mode, so groups of four deliver as many group sums as FP-FP products, and smaller groups deliver more than the pipeline accepts.
Larger groups such as eight also share the pipeline, but they fill only half of its inputs in the MX-FP mode without raising the product rate.
The small group limits truncation, because the truncation grid of each element follows the largest of four exponents, whereas a standard MX block shares one exponent among 32 elements.
Therefore, a group size of four allows the MX-FP mode to share the reduction pipeline with the least truncation among the group sizes that share it, and Section~\ref{sec-arch-numerics} bounds that truncation.

\subsection{Sharing the DSP Array between Modes}
\label{sec-arch-dsp}

The DSP array serves both modes with a single packing scheme, in which each DSP48E2 slice of a group unit multiplies one activation mantissa by the weights of two output lanes, as Figure~\ref{fig:dsp-packing}(a) shows.
Following Section~\ref{sec-back_dsp}, port $B$ receives the shared 9-bit activation mantissa, port $A$ the 5-bit weight of lane 0 and port $D$ the 5-bit weight of lane 1 shifted left by 15 bits.
Because a product of a 9-bit and a 5-bit signed value fits in 14 bits, the 15-bit shift places the two products in disjoint fields of the output $P$, as Figure~\ref{fig:dsp-packing}(b) shows.
A DSP slice packs no third product, because its 5-bit field starts at bit 30 beyond the 27-bit pre-adder.
A narrower field cannot hold a decoded E2M1 element, so its exponent moves into LUT shifts.
Thus, one multiplication serves two lanes at the densest packing that needs no LUT shift, and each lane reads its own field of $P$.

\input{1_Figures/TRETS/architecture/fig-dsp-udpu-float}

Reading each field on its own fails when the product of lane 0 is negative, because this product borrows one from the field of lane 1, as Section~\ref{sec-back_dsp} explains.
The UDPU corrects this borrow inside the DSP slice through port $C$, because the borrow depends only on the operands of lane 0.
It occurs exactly when both operands of lane 0 are nonzero and their signs differ, so the correction logic derives a flag before the multiplication starts.
Registers align this flag with the three-cycle latency of the DSP slice, so port $C$ adds $2^{15}$ in the same operation that produces both products.
Therefore, both products stay exact without an extra pipeline stage, and the packing halves the DSP slices that one product per slice requires.

The FP-FP mode reuses this packing without additional DSP slices, because the packed multipliers process its 5-bit mantissa slices with the same shift, fields of $P$ and sign correction as MXFP4 elements.
The lower mantissa slices of both lanes are mapped to one multiplier and the upper mantissa slices to a second multiplier, and both multipliers receive the same BF16 activation mantissa in port $B$.
Although the lower slice is nonnegative, its product with a negative activation is negative whenever the slice is nonzero and thus subject to the same sign correction.
These partial products are combined in the first adder level of the group unit, which applies the four-bit offset between them, and the MX-FP mode shares this adder level.
In this way, one packing scheme and one sign correction serve both MXFP4$\times$BF16 and BF16$\times$BF16 products, so no mode signal reaches the DSP array and the FP-FP mode adds logic only outside it.

Outside the DSP array, a mode bit travels with every operand, so the UDPU switches modes at instruction boundaries without reconfiguring the fabric or draining the pipeline.
Consequently, a mode switch needs only the idle cycle that already separates consecutive instructions, so the two switches before and after each MLP add no cycle.

\subsection{Numerical Analysis}
\label{sec-arch-numerics}

Sharing the datapath costs accuracy only in the MX-FP mode and mainly through pre-alignment truncation, which this section bounds.
The output error of the UDPU splits into three bounded terms, because only pre-alignment, the reduction pipeline and the final rounding discard bits.
Every product and group sum is exact, as Sections~\ref{sec-arch-compose} and~\ref{sec-arch-dsp} establish.
Pre-alignment truncates toward zero, the reduction pipeline rounds toward minus infinity, and the final rounding to BF16 is to nearest even.
The exact dot product $y = \sum_{i=0}^{n-1} w_i a_i$ of length $n$ combines the BF16 activations $a_i$ with the weights $w_i$, whose values include the shared scale in the MX-FP mode.
For normal activations, finite weights and outputs in the normal BF16 range, the BF16 output $y_{\mathrm{out}}$ satisfies $|y_{\mathrm{out}} - y| \le B_{\mathrm{pre}} + B_{\mathrm{pipe}} + B_{\mathrm{rne}}$ in both modes.
The bound follows from the triangle inequality over two intermediate values, the dot product of the truncated activations and the accumulator value before rounding.
The FP-FP mode skips pre-alignment, so $B_{\mathrm{pre}}$ is zero in that mode.

Pre-alignment truncates each activation $a_i$ toward zero to a multiple $\hat{a}_i$ of $2^{\mathit{exp}_{g(i)}-6}$, the grid step of its group $g(i)$.
This step is the spacing between neighboring values of an aligned activation, so each activation loses at most one step.
An activation whose exponent lies seven or more below $\mathit{exp}_{g(i)}$ becomes zero, because shifting its 7-bit significand by seven or more bits leaves no bit.
Because the weights stay exact, the pre-alignment error $E_{\mathrm{pre}} = \sum_i w_i(\hat{a}_i - a_i)$ of an output is bounded by
\begin{equation}
\label{eq:bpre}
|E_{\mathrm{pre}}| \le B_{\mathrm{pre}} = \sum_{i} |w_i| \min\left(|a_i|, 2^{\mathit{exp}_{g(i)}-6}\right).
\end{equation}

Each activation is truncated only once, so the errors that Equation~\ref{eq:bpre} bounds add without compounding but never cancel for products of one sign, because truncation biases every product toward zero.
Relative to the summed product magnitudes, the largest pre-alignment errors arise when an activation outlier turns small activations with large weights into zeros.
The grid of each group is never coarser than that of the MXINT8 block of 32 elements that contains it~\cite{ocp_mx_spec_v1,rouhani2023microscalingdataformatsdeep}, because MXINT8 takes its grid from the largest exponent of the whole block.
Therefore, the worst-case truncation on this grid stays within twice the rounding bound of the MXINT8 conversion in Equation~\ref{eq:mx_quant}, because truncation loses up to one grid step where rounding loses half.
In exchange, pre-alignment needs no rounding logic per activation.

The pipeline and rounding terms are not a cost of sharing in the FP-FP mode, because the BF16 DPU of Section~\ref{sec-back_dpu} incurs both with the same reduction pipeline.
The rounding term $B_{\mathrm{rne}} = 2^{-8}|y_{\mathrm{out}}|/(1-2^{-8})$ covers half a unit in the last place of the BF16 output, whose significand holds 8 bits.
The pipeline term $B_{\mathrm{pipe}}$ of the FP-FP mode is smaller, because the accumulator keeps 23 fraction bits and loses under four units in their last place per cycle, about $2^{-21}$ of the summed magnitudes.
This loss reaches one BF16 rounding only after about 8,000 cycles, or half a million FP-FP products per lane, far beyond the dot-product length of the models that Section~\ref{sec:eval} evaluates.
In the MX-FP mode, each group sum takes its exponent from its group exponent and shared scale rather than its magnitude, so a group with a large activation and small weights lowers accumulation precision.
The smallest nonzero product on the grid of that group is one grid step times the smallest nonzero weight of its block.
Still, the pipeline then loses only a small fraction of that product per cycle, whereas truncation can lose the whole product or more on every activation.

Therefore, the accuracy cost of sharing in the MX-FP mode is mainly a truncation of at most one grid step per activation, on a grid never coarser than that of a standard MXINT8 block.
Groups of four also keep the truncation that an activation outlier causes within three neighbors, so this cost stays local to each group.
Section~\ref{sec-eval-accuracy} checks the RTL outputs on real inputs against this bound and measures the effect of the truncation on model accuracy.

%% file: 1_Figures/TRETS/architecture/fig-arch-udpu-float.tex
\begin{figure}[!t]
\centering
\includegraphics[width=\linewidth]{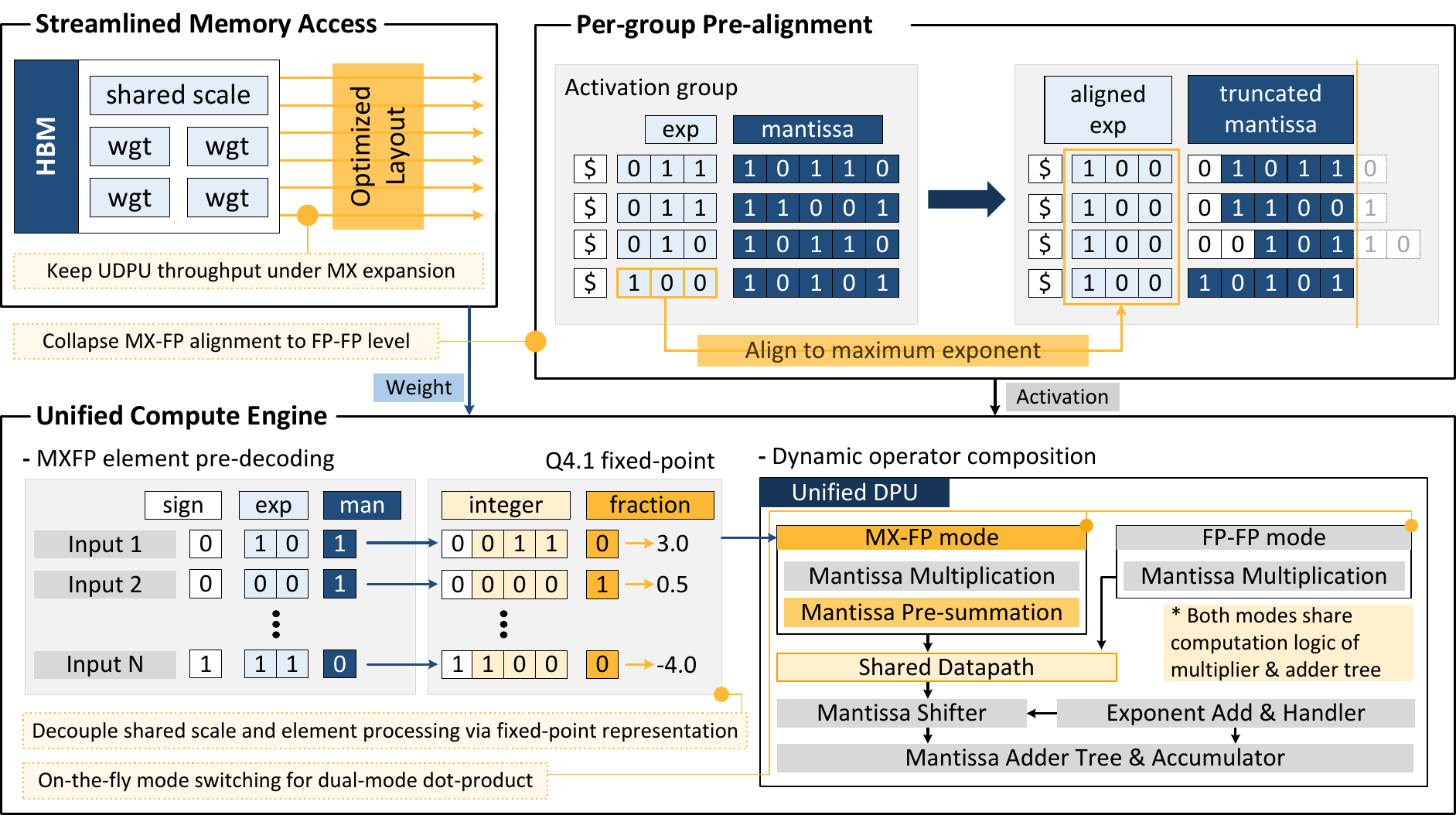}
\Description{Block diagram of the compute engine, in which pre-aligned activations and decoded weights feed dot-product units with shared multipliers and adder trees.}
\caption{Overall \sysname compute engine. Activations are pre-aligned in groups of four and MXFP4 weights are decoded to fixed point, so the UDPU runs both modes on shared multipliers and one reduction pipeline.}
\label{fig:arch-udpu}
\end{figure}

%% file: 1_Figures/TRETS/architecture/fig-udpu-detail-float.tex
\begin{figure}[!t]
\centering
\includegraphics[width=\linewidth]{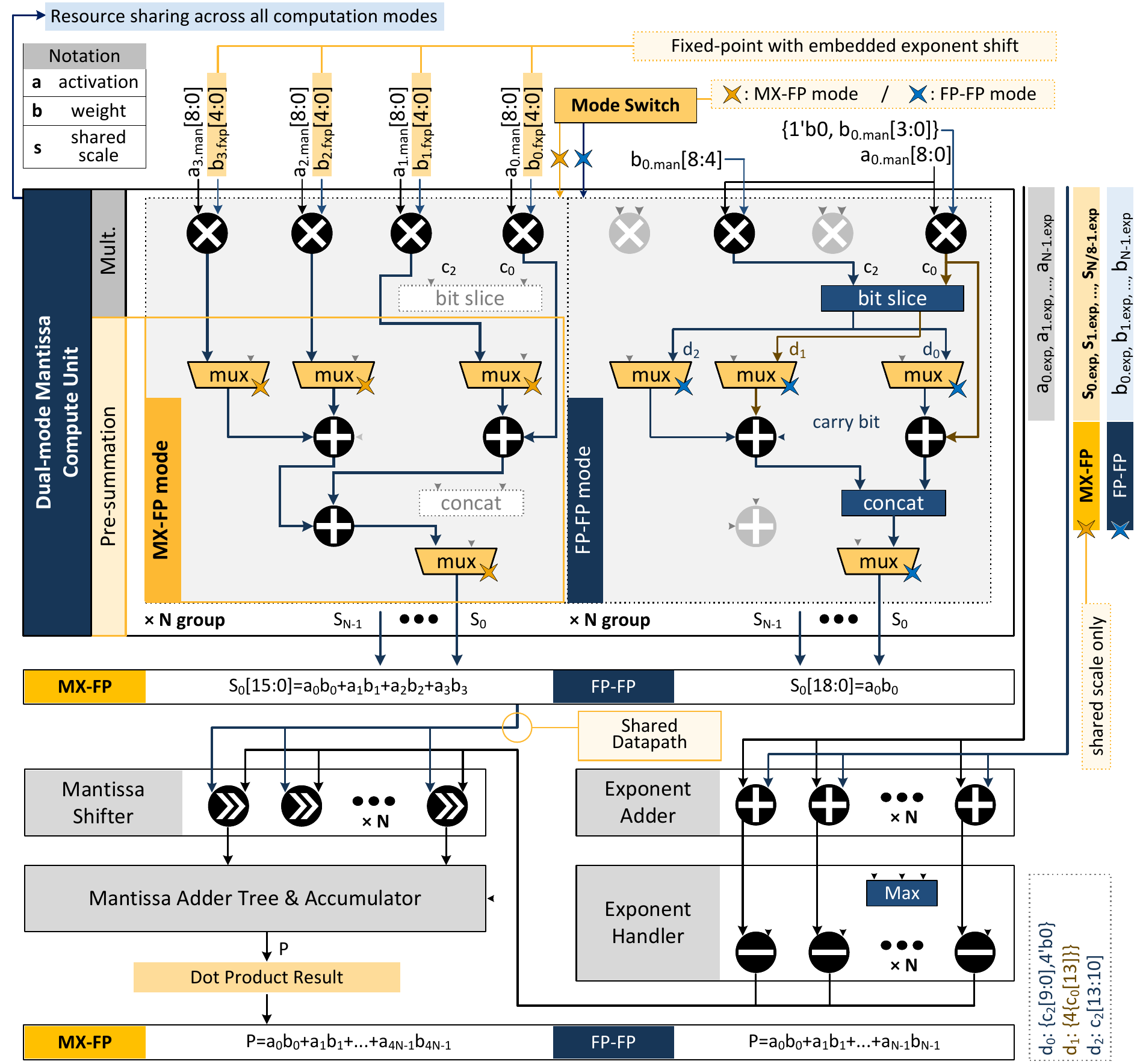}
\Description{Microarchitecture of the unified dot-product unit, from the operand path through the DSP product stage and the reduction pipeline to the accumulator.}
\caption{Microarchitecture of the UDPU. In the MX-FP mode, the activation exponents are indexed by four-element group and the shared scales by 32-element block.}
\label{fig:udpu-detail}
\end{figure}

%% file: 1_Figures/TRETS/architecture/alg-prealign.tex
\begingroup
\setlength{\textfloatsep}{4pt}
\begin{algorithm}[t]
\footnotesize
\caption{Pre-alignment of one group of BF16 activations in the MX-FP mode}
\label{alg:prealign}

{\raggedright
\textbf{Input:} BF16 activations $a_i$ for $i \in [0, G-1]$, each with $\mathit{sign}_i$, unbiased $\mathit{exp}_i$ and 7-bit $\mathit{man}_i$ \\
\textbf{Output:} group exponent $\mathit{exp}_g$ and 8-bit two's complement $\mathit{element}_i$, where $\mathit{element}_i \cdot 2^{\mathit{exp}_g-6}$ is $a_i$ truncated toward zero \\
\textbf{Parameter:} $G=4$ (group size) \par
}
\vspace{-0.6mm}

\begin{algorithmic}[1]
\State $\mathit{exp}_g \gets \max_{i}(\mathit{exp}_i)$ \Comment{largest exponent of the group}
\For{$i \gets 0$ to $G-1$ \textbf{in parallel}}
  \State $\mathit{significand}_i \gets \{1,\, \mathit{man}_i[6{:}1]\}$ \Comment{7-bit integer, hidden bit and top six bits}
  \State $\mathit{aligned}_i \gets \mathit{significand}_i \gg (\mathit{exp}_g - \mathit{exp}_i)$ \Comment{shifted-out bits dropped}
  \State $\mathit{element}_i \gets (-1)^{\mathit{sign}_i} \cdot \mathit{aligned}_i$ \Comment{stored as 8-bit two's complement}
\EndFor
\end{algorithmic}
\end{algorithm}
\endgroup

%% file: 1_Figures/TRETS/architecture/fig-dsp-udpu-float.tex
\begin{figure}[!t]
\centering
\includegraphics[width=\linewidth]{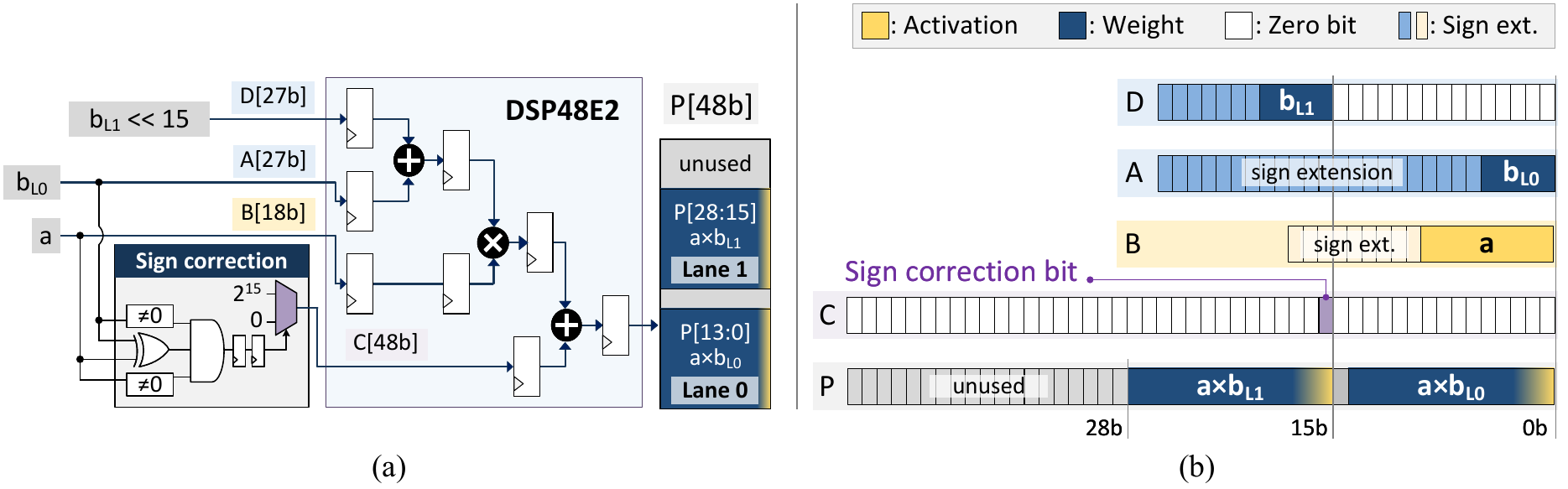}
\Description{Diagram of one DSP48E2 slice that multiplies one activation mantissa by two packed weights with a sign correction, and the bit fields of its ports and output.}
\caption{DSP packing in one DSP48E2 slice of a group unit. (a) Two lane weights share one activation mantissa, and port $C$ adds the sign correction. (b) Bit fields of the ports and of $P$.}
\label{fig:dsp-packing}
\end{figure}

%% file: 0_Text/5_top_architecture.tex
\section{Platform Integration}
\label{sec-platform}

\sysname integrates the UDPU into an FPGA platform for LLM inference that runs complete models end to end, which shows that the unit is feasible within a deployed accelerator.
Section~\ref{sec-platform-org} outlines the organization of the platform, and Section~\ref{sec-platform-mx} describes the MX operand delivery of \sysname, which supplies operands in the order that the UDPU consumes them.

\subsection{Platform Organization}
\label{sec-platform-org}

The platform executes each decoder layer of an LLM on a compute engine and a vector processing unit (VPU), which exchange activations through a register file, as Figure~\ref{fig:platform}(a) shows.
The compute engine of Section~\ref{sec-arch} computes every matrix-vector product, whereas the VPU computes vector operations such as normalization and softmax.
The streamlined memory access (SMA) unit fetches weights and the KV cache from HBM, and the operand collector feeds them to the UDPUs together with the corresponding activations.

\input{1_Figures/TRETS/architecture/fig-overall-arch-float}

A decoder layer runs on these units in the sequence of Figure~\ref{fig-mode-share}(a).
The compute engine executes the projections, the attention products and the MLP, each in the mode of its operands.
The VPU runs its vector operations alongside these products and overlaps the compute engine within a phase.
The platform runs single-batch decoding and prefills a prompt token by token, so a prompt token at position $t$ costs about as much as a decode token at context length $t$.

\subsection{Delivering MX Operands to the UDPU}
\label{sec-platform-mx}

The operand collector hosts the operand path of Section~\ref{sec-arch-compose} and issues both operands of every product to the UDPUs in the same cycle.
Activations from the register file pass the pre-alignment of Section~\ref{sec-arch-prealign} in the MX-FP mode, and the operand collector keeps each vector in one buffer that all lanes share.
In contrast, weights and the KV cache reach each lane on a stream of its own, which joins the 512-bit lines of two HBM AXI channels into one 1,024-bit word per cycle.
A word holds 64 BF16 operands in the FP-FP mode and 256 MXFP4 elements of eight blocks in the MX-FP mode, so each word supplies one cycle of products for its lane in either mode.
The 32 channels thus deliver 2,048 bytes per clock cycle, enough for 3,855 MXFP4 weights at 4.25 bits each or 1,024 BF16 weights or KV elements.
At 4,096 MX-FP or 1,024 FP-FP products per cycle, the eight UDPUs match this bandwidth in the FP-FP mode and exceed it in the MX-FP mode.

\sysname carries the shared scales on the weight stream itself, so the weights of each lane and their scales stream from consecutive HBM addresses without a second address stream.
The offline weight mapper of \sysname places one word of up to 128 scales ahead of the element words that hold their blocks, as Figure~\ref{fig:platform}(b) shows for one channel of a lane.
A 1,024-bit word holds exactly 128 eight-bit scales, so every element word stays aligned to whole blocks, whereas a layout with a scale beside each block splits blocks across words.
Because each element word holds eight blocks, the 128 scales of one scale word cover the blocks of up to 16 element words.

The exponent adders of Section~\ref{sec-arch-scale} take one scale per group of four elements, whereas the stream delivers 128 scales per word ahead of their blocks.
A small buffer in the operand collector retains the current scale word of each lane and supplies eight of its scales with every element word, one for each block of that word.
Each scale is repeated for the eight groups in its block, so every group of four elements receives a copy of its block scale.
Thus, the operand collector expands every scale word into one scale per group, so each lane consumes scales and elements in stream order with a buffer of only 128 scales.
The FP-FP mode skips this expansion, because its stream carries no scale words and the exponent adders receive the exponent of each weight-side BF16 operand instead.

%% file: 1_Figures/TRETS/architecture/fig-overall-arch-float.tex
\begin{figure}[!t]
\centering
\includegraphics[width=\linewidth]{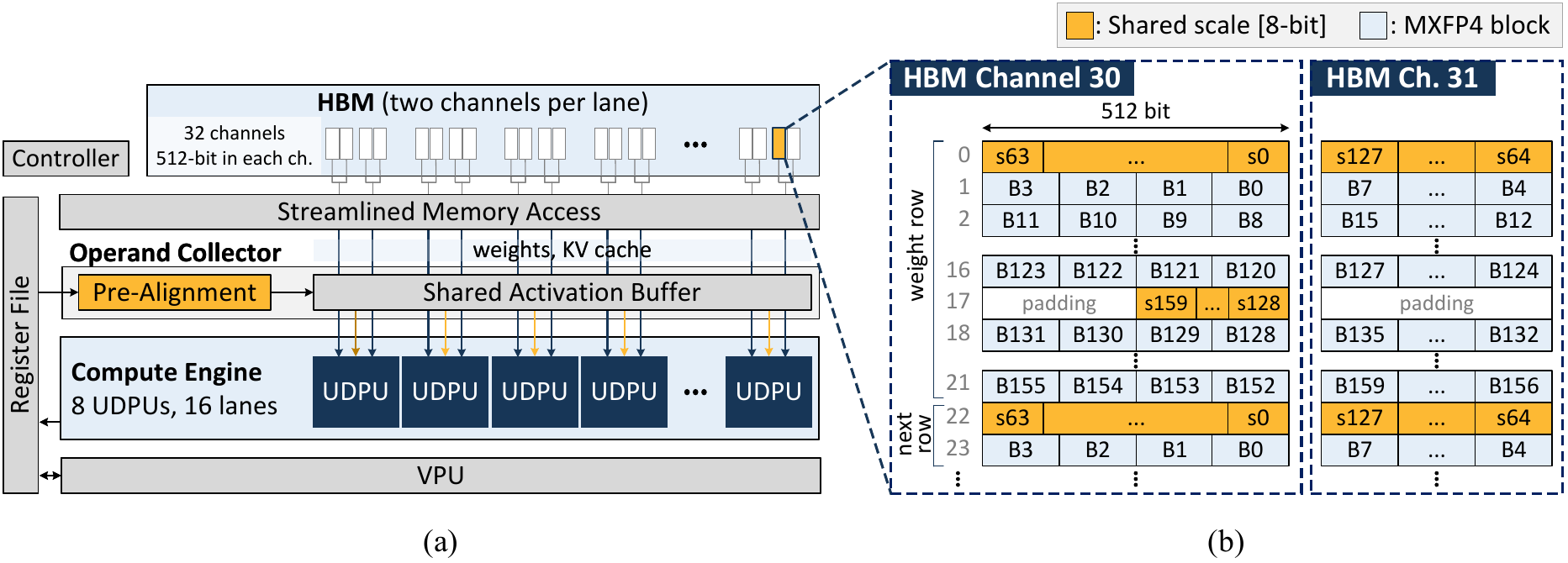}
\Description{Block diagram of the platform around the compute engine with its HBM channels, and the layout of scale lines and element lines in one HBM channel.}
\caption{(a) The platform around the \sysname compute engine. (b) Lines of HBM channel 30 in the MX-FP mode, where a scale line precedes up to 16 element lines.}
\label{fig:platform}
\end{figure}

%% file: 0_Text/6_implementation.tex
\section{FPGA Implementation}
\label{sec-impl}

\sysname is implemented and evaluated on one AMD Alveo U55C board, whose HBM and three super logic regions (SLRs) determine where the eight UDPUs of Section~\ref{sec-platform-mx} reside.
The bitstream runs its RTL kernels at 185.4 MHz and reaches 16 GB of HBM through 32 AXI channels with a peak bandwidth of 460 GB/s.
Vitis 2022.2 builds the bitstream with the AggressiveExplore directive for routing and for post-route physical optimization.

\subsection{Floorplan and SLR Mapping}
\label{sec-impl-floorplan}

The bitstream divides the 16 UDPU lanes among three RTL kernels, one in each SLR, as Figure~\ref{fig-floorplan} shows.
The kernel in SLR0 is adjacent to the HBM stacks, so it connects to all 32 HBM ports and contains four of the lanes.
The kernel in SLR1 contains eight lanes, and the kernel in SLR2 contains the other four together with the VPU, the controller and the host interface.

\input{1_Figures/TRETS/implementation/fig-floorplan-float}

The lanes are spread over all three SLRs so that each SLR contributes part of the logic, which balances the resource use.
Because all 32 HBM ports reside in SLR0, the kernel there forwards the weight streams of the other 12 lanes, so 24 streams of 512 bits cross into SLR1 every cycle.
Eight of these streams continue into SLR2 to feed its four lanes.
Thus, the balanced distribution trades these SLR crossings for lower routing congestion within each SLR.

MXFP4 weights occupy the HBM in the word order of Section~\ref{sec-platform-mx}, so each matrix streams from consecutive addresses together with its scales.
Because a scale word serves at most 16 element words, the scales add at least 6.25\% to the weight traffic, the fraction that their eight bits add to a block of 32 four-bit elements.

\subsection{Resource Breakdown}
\label{sec-impl-resource}

The table in Figure~\ref{fig-floorplan} breaks the post-route resources of the three kernels down by module and compares them with the capacity of the U55C.
The eight UDPUs are the largest module, with 17.7\% of the LUTs and 24.1\% of the DSP slices of the device.
Each UDPU serves two lanes and spends 128 multiplier DSP slices on each of them.
The 2,176 DSP slices of the UDPUs comprise 2,048 multipliers and 128 accumulators, eight per lane.
Together they hold 94\% of the DSP slices and 53\% of the LUTs that the kernels use, so the rest of the accelerator relies mainly on LUTs and memory.
Section~\ref{sec-eval-resource} separates the part of these UDPU resources that the FP-FP mode adds.

The UDPUs use only nine BRAM tiles, because weights, the KV cache and their scales stream from HBM instead of on-chip buffers.
Most of the 915 BRAM tiles and 44\% of the FFs belong to the row labeled Etc., which collects the remaining logic of the three kernels.
This logic consists mainly of the operand collector and the register slices at the SLR crossings, which ease timing closure on the paths between the SLRs.
The lane count stays at 16 despite the free resources, because each lane takes the 1,024-bit word of two AXI channels per cycle and the 32 channels feed no more than 16 lanes.

%% file: 1_Figures/TRETS/implementation/fig-floorplan-float.tex
\begin{figure}[!t]
\centering
\includegraphics[width=0.8\linewidth]{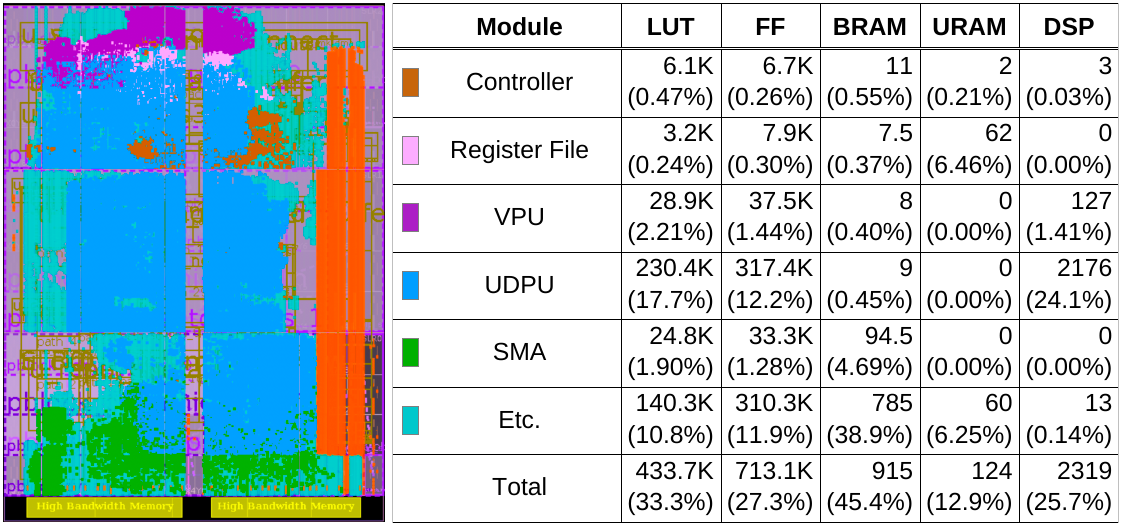}
\Description{Post-route floorplan of the U55C device with the placed cells of each module in a different color and a list of the resources of each module.}
\caption{Post-route floorplan of the deployed bitstream on the U55C with the resources and device proportion of each module. Orange cells on the right belong to the static shell.}
\label{fig-floorplan}
\end{figure}

%% file: 0_Text/7_evaluation.tex
\section{Evaluation}
\label{sec:eval}

This section quantifies what the FP-FP mode and the DSP packing cost within the UDPU and compares the UDPU with alternative designs on the same fabric.
Next, it tests the accuracy of pre-alignment and measures end-to-end decoding on the board against a GPU and FPGA accelerators.
Finally, it discusses the limitations of these results.

\subsection{Experimental Setup}
\label{sec-eval-setup}

The board measures only the latency of whole tokens, and its bitstream contains only the UDPU, so it shows neither the phases inside a token nor the designs that the UDPU replaces.
Therefore, we add RTL simulation for the phases and out-of-context place-and-route for the UDPU variants and the alternative designs of Section~\ref{sec:split-unified}.
A software emulator of the datapath arithmetic adds the accuracy against exact products.
Section~\ref{sec-eval-e2e} reports the board runs, Sections~\ref{sec:mode-share} and~\ref{sec:discussion} use the simulated phases, and Sections~\ref{sec-eval-resource} and~\ref{sec-eval-alternatives} report the unit runs.
Section~\ref{sec-eval-accuracy} uses the emulator and RTL simulation of the units, and we state the source of every number so that simulated values are never mistaken for board measurements.

\textbf{Board measurement.}
Every board measurement runs on one U55C board, where the platform of Section~\ref{sec-platform-org} runs single-batch decoding and prefills a prompt token by token.
Table~\ref{tab-eval-setup} lists the board next to the GPU baseline.
\input{1_Figures/TRETS/evaluation/tab-eval-setup}
Each run prefills a prompt of $P$ tokens and decodes $D = 256$ tokens for one request, at contexts $C = P + D$ from 512 to 16K.
We report the time to first token (TTFT) and the time per output token (TPOT), the mean gap between the decode tokens, from host timestamps on every streamed token.
Therefore, both metrics include the host runtime.
Power is the mean card power that \texttt{xbutil} reports over the decode window, without the host.
Each configuration runs one warm-up and five measured repetitions.
The 95\% confidence interval of every mean lies within 0.4\% for TPOT and TTFT and within 1.2\% for power and energy.
All board runs use the same bitstream.

\textbf{GPU baseline.}
The A40 runs the vLLM serving engine~\cite{kwon2023efficient} with Marlin~\cite{frantar2025marlin}, a mixed-precision kernel optimized for memory-bound decoding on Ampere GPUs, which gives the GPU a strong MXFP4 baseline.
This engine extends Marlin to MXFP4, so the A40 keeps the MXFP4 weights of the board packed in device memory at 4.25 bits per weight and converts them to BF16 in registers just before the tensor-core products.
Therefore, the A40 reads the same weight bytes per token as the board, so the gap between them comes from the platform and not from the weight format.
Its power is the card power that \texttt{nvidia-smi} reports over the same window.
It also runs one warm-up and five measured repetitions, with 95\% confidence intervals within 1.5\% for TPOT and 2.9\% for power and energy.
We compare decoding only, because only there do both platforms read the same bytes per token, whereas the GPU prefills a whole prompt as compute-bound matrix products that read each weight once per prompt.

\textbf{RTL simulation.}
The phase proportions come from VCS simulations of the RTL of the deployed bitstream, which run one output token at chosen positions through one and two decoder layers and then through the LM head and sampling.
We repeat these layers to the full depth of each model and interpolate between the simulated positions, so the proportions count RTL cycles and not board time.

\textbf{Unit place-and-route.}
We place and route every unit of Sections~\ref{sec-eval-resource} and~\ref{sec-eval-alternatives} out of context with the build flow of Section~\ref{sec-impl}.
Every unit closes timing at about 200~MHz, above the 185.4~MHz of the board, so equal products per cycle also mean equal products per second.

\textbf{Accuracy evaluation.}
The output error of Section~\ref{sec-eval-accuracy} comes from RTL simulation of the UDPU and the FIGNA- and FIGLUT-style units of Section~\ref{sec-eval-alternatives}.
Each design computes 122,880 outputs from the gate, up, and down projections of five dense models.
The error of an output is its difference from the exact dot product computed in FP64, relative to the sum of absolute products.
For the model accuracy, a PyTorch emulator applies the bit-exact alignment of each of these designs and accumulates in FP32, and the exact baseline skips the alignment.
On these dot products, this alignment is bit-exact with the RTL, whereas the vector operations such as normalization and softmax run as standard PyTorch operations.
The emulator measures WikiText-2 perplexity over non-overlapping 1,024-token windows and five-shot ARC-Challenge~\cite{clark2018think} and MMLU~\cite{hendrycks2020measuring} in lm-eval on the instruct releases of six models.
GPT-OSS-20B is the exception, with UltraChat transcripts and zero-shot questions in its chat format.
The dense models take MXFP4 MLP weights from GPTQ~\cite{frantar2022gptq}, whereas GPT-OSS-20B keeps its released MXFP4 weights.

\subsection{Resource Cost of the Unified Datapath}
\label{sec-eval-resource}

A unified datapath is justified only if its FP-FP mode requires few resources beyond the MX-FP mode, so we quantify this overhead on the UDPU and the saving of the shared DSP packing.
We isolate the cost of each feature with ablation variants that change only that feature of the UDPU and pass the place-and-route flow of Section~\ref{sec-eval-setup}.
The MX-only variant keeps only the logic of the UDPU that the MX-FP mode uses.
Two variants without the packing of Section~\ref{sec-arch-dsp} keep the product rate of the UDPU and compute one product per DSP slice or in LUT multipliers.
Each variant also includes the pre-alignment and BF16 activation buffering, because the two modes need them even though the board places them in the operand collector outside the UDPU row of Figure~\ref{fig-floorplan}.
Table~\ref{tab-eval-variants} lists the resources per lane as half of each variant, which holds one UDPU with its two lanes.

\input{1_Figures/TRETS/evaluation/tab-eval-variants}
\input{1_Figures/TRETS/evaluation/tab-eval-stages}

The FP-FP mode adds no DSP slices to the UDPU and only about 21\% of the LUTs that a separate BF16 lane adds.
Its products need no DSP slices of their own, because Section~\ref{sec-arch-dsp} maps its mantissa slices onto the packed multipliers of the MX-FP mode.
It adds 2,677 LUTs per lane, which are 14.9\% of the UDPU lane and 17.5\% more than the MX-only variant.
Against an MX-FP unit designed without the FP-FP mode, this increase can be larger, because the MX-only variant keeps the pipeline depths and bit widths of the UDPU.
The BF16 lane that the bandwidth-matched split of Section~\ref{sec-eval-alternatives} adds at the same FP-FP rate takes about 12,600 LUTs and 40 DSP slices.

The overhead of the FP-FP mode stays small because both modes share the reduction pipeline, which takes half of the LUTs of a lane in Table~\ref{tab-eval-stages}.
Its alignment is the largest LUT consumer at 41\%, so a split datapath of Section~\ref{sec:split-unified} duplicates it in its separate BF16 lane.
The weight decoding and multipliers hold all 128 multiplier DSP slices with 34\% of the LUTs.
The FP-FP mode adds about 90\% of its LUTs to the BF16 activation buffering and to the weight decoding and multipliers.
In the second of these steps, the FP-FP logic decodes BF16 weights into the two 5-bit mantissa slices of Section~\ref{sec-arch-compose} and recombines the two partial products of each BF16 product.
In contrast, the shared reduction pipeline grows by fewer than 200 LUTs, because both modes feed it signed mantissas with their own exponents in the same format.

Every part of the lane works in at least one mode.
In the MX-FP mode, the lane uses all of its DSP slices but only 85\% of its LUTs, the part that the MX-only variant keeps.
In the FP-FP mode, the lane leaves the pre-alignment and the MXFP4 decoders idle and uses 64 of the 128 multiplier DSP slices.
Half of the slices suffice for this mode, because it needs only about a quarter of the MX-FP product rate and each BF16 product occupies two multipliers.

Whereas the FP-FP mode costs LUTs, the packing saves DSP slices at a small LUT cost.
It halves the multiplier DSP slices of a lane from 256 to 128 and costs only 2.0\% of the LUTs of a UDPU lane.
The packed slices keep the three-cycle latency of slices without packing, so the sign correction of Section~\ref{sec-arch-dsp} adds registers but no cycle of latency.
Mapping the multipliers to LUTs removes all 128 multiplier DSP slices of a lane but needs 58.8\% more LUTs than the UDPU.
This variant uses generic 9-by-5-bit multipliers, because the FP-FP mode feeds arbitrary 5-bit mantissa slices to the same multipliers.
Over the 16 lanes, the packing saves 2,048 DSP slices or 22.7\% of the device for 0.4\% of its LUTs, whereas LUT multipliers in their place take another 13.0\% of its LUTs.

Therefore, the UDPU supports both modes with few resources beyond those of the MX-FP mode.
Both modes run on one array of 128 multiplier DSP slices per lane, half of what one product per slice requires, and the FP-FP mode adds 17.5\% to the LUTs of an MX-only lane.
Section~\ref{sec-eval-alternatives} weighs this cost against the DSP slices that alternative designs need at the same throughput.

\subsection{Unit-Level Comparison with Alternative Designs}
\label{sec-eval-alternatives}

To test whether the UDPU serves both modes with fewer resources than the alternatives of Section~\ref{sec:split-unified}, we route each as a unit at the same per-mode throughput.
The bandwidth-matched split sizes each of its two lanes to the product rate of a UDPU lane in its mode, which keeps pace with HBM.
Its MX-FP lane is the MX-only variant of Section~\ref{sec-eval-resource}, and its BF16 lane is the BF16 DPU of Section~\ref{sec-back_dpu}.
The FP-only extension dequantizes the MXFP4 weights in front of the same BF16 lanes.
Two further splits replace the MX-FP lane with FPGA reimplementations of FIGNA~\cite{jang2024figna} and FIGLUT~\cite{park2025figlut}, the integer and lookup-table methods of Figure~\ref{fig-dpu}(b).
Both keep the parameters of the original designs where MXFP4 weights and single-batch decoding allow.
They target the FP-INT product only, so each split pairs its reimplementation with the BF16 lane that the FP-FP mode needs, and Table~\ref{tab-eval-alternatives} also lists the MX-FP lane of each reimplementation alone.
Both originals are 28-nm ASIC designs that save area on multipliers~\cite{jang2024figna,park2025figlut}, so their reimplementations test whether this saving survives on an FPGA, where multipliers are hard DSP slices.
Thus, these results hold for the reimplementations on this fabric, not for the original ASIC designs.

\input{1_Figures/TRETS/evaluation/tab-eval-alternatives}

The designs differ only in their datapaths.
Every BF16 lane and the FIGNA-style MX-FP lane pack two products into each DSP slice as the UDPU does~\cite{fu2017wp486}, so differences in DSP slices reflect the datapath rather than its mapping.
Every alternative is bit-exact with the UDPU in the FP-FP mode, and the bandwidth-matched split is also bit-exact with it in the MX-FP mode.
Table~\ref{tab-eval-alternatives} lists the post-route resources of each design at the throughput of one UDPU lane, 256 MX-FP and at least 64 FP-FP products per cycle.
The FIGLUT-style unit keeps the array of the original design, so it computes 3.2$\times$ this MX-FP rate.
Therefore, its row divides the MX-FP resources of this unit by 3.2 and adds the packed BF16 lane of the bandwidth-matched split.

Under the sequential schedule, the bandwidth-matched split needs 1.29$\times$ the DSP slices of the UDPU and at most 1.55$\times$ its LUTs, where the DSP count includes the accumulators.
Its extra LUTs come from the BF16 lane, which repeats the alignment and summation that the two modes of the UDPU share.
This cost does not depend on the proportion of time in the FP-FP mode, because each phase runs on one lane alone.
Therefore, the UDPU needs 23\% fewer DSP slices and 36\% fewer LUTs than the bandwidth-matched split at the same per-mode throughput.

The FP-only extension computes 64 products per cycle in either mode, so matching a UDPU lane takes four of its lanes in the MX-FP mode but one in the FP-FP mode.
At the MX-FP rate, these four lanes need 3.2$\times$ the LUTs of a UDPU lane.
Most of these LUTs align and sum products in the four BF16 lanes, whereas the dequantizers that convert MXFP4 weights to BF16 take only about 10\% of them.
At the FP-FP rate, a UDPU lane needs 1.26$\times$ the LUTs of one FP-only lane and 96 more DSP slices for four times the MX-FP rate.
Therefore, the FP-only extension keeps pace with MXFP4 weights only by replicating its BF16 lane four times, so its LUTs rather than its DSP slices limit how many lanes fit on the device.

The FIGNA-style split keeps its multipliers in DSP slices but spends LUTs on a wider and more accurate alignment, so it needs 2.5$\times$ the LUTs of the UDPU.
Its MX-FP lane alone already needs more LUTs than the whole UDPU.
Sharing the activation logic of both designs over eight lanes as in the kernels lowers the split's LUT multiple only to about 2.3$\times$, which still excludes its 55\% wider activation buffer outside the unit.
It aligns its activations at 30 bits, the width that the original design sets for FP32 accumulation, so its integer sums grow wide as well.
Its multipliers take each aligned activation as a pair of 7-bit chunks~\cite{jang2024figna}, which narrows ASIC multipliers but on the FPGA adds LUT shifters that realign each chunk product.
The chunk shifts and integer sums after the multipliers take about half of its MX-FP LUTs and the alignment before them over a quarter.
In contrast, the UDPU pre-aligns each group of four activations at 7 bits, so each product fits one DSP field without chunks and its sums stay at the FP-FP width.

The FIGLUT-style split shows that the saving of table lookup on an ASIC turns into a LUT cost on the FPGA.
It needs 0.44$\times$ the DSP slices of the UDPU but 3.6$\times$ its LUTs, and most of these LUTs belong to its MX-FP part.
Each table is indexed by the same bit of several weights, so every weight bit plane takes its own pass of table reads.
MXFP4 weights need five planes because twice an E2M1 value is a 5-bit integer, so the same product rate takes five times the table reads of one plane.
The read-and-accumulate units that perform these reads and the FP32 accumulators take two thirds of its MX-FP LUTs, whereas the logic that our reimplementation adds to the original design takes only 9\% of them.
Even without its FP32 accumulators and with only the four planes of the original 4-bit weights, this part still needs about 1.7$\times$ the LUTs of the UDPU.
Unlike these bit-serial table reads, the UDPU multiplies all five bits of an element in one pass through a packed DSP slice.

In summary, under the sequential schedule, the UDPU reaches both per-mode rates with at most 65\% of the LUTs of any alternative and with fewer DSP slices than every alternative that multiplies in DSP slices.
Because LUTs are exhausted first on the FPGA, no alternative keeps pace with MXFP4 weights within the resources of 16 UDPU lanes, as the right columns of Table~\ref{tab-eval-alternatives} show.
The UDPU keeps this lead because its two modes share one reduction pipeline and one set of packed DSP slices, which every alternative either duplicates or replaces with LUTs.
This advantage comes with a 7-bit alignment against the 30 and 15 bits of the FIGNA- and FIGLUT-style designs, so Section~\ref{sec-eval-accuracy} measures what pre-alignment costs in accuracy against exact products.

\subsection{Accuracy Comparison of Pre-Alignment and Alternative Designs}
\label{sec-eval-accuracy}

Pre-alignment truncates the activations to share the reduction pipeline, so its effect on accuracy must be measured both in each dot-product output and in the whole model.
Each BF16 activation keeps a 7-bit significand under the shared exponent of its group of four, as Section~\ref{sec-arch-prealign} describes.
This truncation is the only change that the MX-FP mode makes to an operand.
Attention runs in the FP-FP mode, so pre-alignment acts only on the MLP inputs and its error does not grow with the context length.
The FIGNA- and FIGLUT-style units align 30 and 15 bits in their MX-FP lanes instead of 7 and are bit-exact with the UDPU everywhere else.
Therefore, we compare the UDPU and these two units with exact MXFP4$\times$BF16 products at both levels, with the RTL simulation and the emulator of Section~\ref{sec-eval-setup}.

\input{1_Figures/TRETS/evaluation/fig-eval-unit-error-float}

At the output level, pre-alignment adds an error that stays within the BF16 rounding for almost every output, as Figure~\ref{fig-eval-unit-error} shows.
Every design writes BF16 outputs, so even an exact unit has an error of up to $2^{-8}$ of the sum of absolute products $\sum |a w|$ through rounding.
The FIGNA- and FIGLUT-style units keep all of their outputs within this rounding because of their wide alignment.
At a group size of four, the UDPU keeps 99.99\% of its outputs within the same rounding.
Even its largest error reaches only 27\% of the bound $B_{\mathrm{pre}} + B_{\mathrm{pipe}} + B_{\mathrm{rne}}$ of Section~\ref{sec-arch-numerics}.
However, its error has about five times the standard deviation of the FIGNA- and FIGLUT-style units.
Its error is also biased, because truncation always shrinks an activation and moves the outputs toward zero by 0.015\% of the sum of absolute products on average.
Rounding to nearest avoids this bias, but it takes 40\% more LUTs in the quantizer, as Section~\ref{sec-arch-prealign} explains.
Therefore, the output error of the UDPU is larger than that of the alternatives but stays within the BF16 rounding in almost every output.

At the model level, the larger output error of pre-alignment has only a small effect on accuracy, as Table~\ref{tab-eval-accuracy} lists.
For reference, the MXFP4 weights alone raise perplexity by 1.2\% to 10\% over BF16 weights.
Against this loss, pre-alignment at a group size of four adds at most 0.7\% on the dense models.
The FIGNA- and FIGLUT-style alignment stays within 0.03\% of exact products, so this 0.7\% is also the gap between the UDPU and the alternatives.
On ARC-Challenge and MMLU, the MXFP4 weights lose up to 2.7 points, whereas every design stays within 0.6 points of exact products on the dense models.
On GPT-OSS-20B, no design raises perplexity or lowers the ARC-Challenge accuracy, and MMLU stays within 0.4 points of exact products.
Therefore, the larger output error of the UDPU changes perplexity by at most 0.7\% and task accuracy by at most 0.6 points against exact products on the dense models.

\input{1_Figures/TRETS/evaluation/tab-eval-accuracy}

Across group sizes, larger groups raise the output error and smaller groups gain little accuracy, which supports the group size of four.
A larger group truncates more activations to the grid of its largest exponent, so the standard deviation of the output error in Figure~\ref{fig-eval-unit-error} grows by about 1.4$\times$ with every doubling.
The mean error toward zero also grows from 0.015\% at a group size of four to 0.030\% at 32.
Perplexity in Table~\ref{tab-eval-accuracy} stays within $+0.8\%$ of exact products up to a group size of eight on every dense model, and only Qwen3-1.7B passes 1\% at 16 and 32.
ARC-Challenge and MMLU stay within 0.6 points of exact products at every group size on the dense models.
Smaller groups gain little, because a group size of one shares no exponent but still causes at least half of the perplexity change at four.
As Section~\ref{sec-arch-prealign} explains, smaller groups deliver more sums than the reduction pipeline of the FP-FP mode accepts, and larger groups leave it partly idle.
Therefore, the hardware sets the group size to four, which also has the smallest output error among the sizes that fit the pipeline.

In summary, the UDPU has a larger output error than the alternatives but stays within 0.7\% of their perplexity, and at this cost it saves the LUTs and DSP slices that Section~\ref{sec-eval-alternatives} reports.

\subsection{End-to-End Performance and Energy Analysis}
\label{sec-eval-e2e}

Sections~\ref{sec-eval-resource} to~\ref{sec-eval-accuracy} evaluate the UDPU as a unit, whereas a decode token interleaves the two modes and also spends time outside the dot-product units.
Therefore, this subsection asks how much of the byte saving of MXFP4 weights reaches the end-to-end latency of a real board.
We measure this gain in the TPOT and the TTFT across contexts and model sizes, and we compare the board with a GPU and with FPGA accelerators.
Figure~\ref{fig-eval-board} reports the board runs of Section~\ref{sec-eval-setup} for six models and contexts from 512 to 16K, where three BF16 configurations exceed the HBM and appear as out-of-memory (OOM) slots.
We measure GPT-OSS-20B in MXFP4 only, because it is an MXFP4-native model whose experts are released in this format.

\input{1_Figures/TRETS/evaluation/fig-eval-board-float}

\textbf{Speedup from MXFP4 weights.}
At a context of 512, MXFP4 weights shorten the TPOT by an average of 1.54$\times$, with a maximum of 1.87$\times$ for Llama-3.1-8B.
An MXFP4 weight takes 4.25 bits with its part of the block scale, so it reads 3.76$\times$ fewer bytes than a BF16 weight.
Yet per decode token, the BF16 model reads only 1.79$\times$-2.19$\times$ the bytes of the MXFP4 model, because only the MLP weights change format.
The speedup stays below even this byte ratio, because each token also spends a fixed time on work that streams no weights, such as sampling and the element-wise VPU operations.
We divide each TPOT into this fixed time and the time to stream the bytes of the token, and we assume that both formats share the fixed time and the stream rate.
The BF16 and MXFP4 runs of each model and context then give a stream rate of 349-358~GB/s for every pair, about 93\% of the 380~GB/s that the 32 HBM ports of the kernels carry.
At a context of 512, the fixed time ranges from 5.0~ms for Llama-3.2-1B to 14.8~ms for Qwen3-8B, or 27\%-67\% of the MXFP4 TPOT.
Part of this fixed time comes from the VPU, because both formats run on the MX build and its VPU is half as wide as that of a BF16-only build, as Section~\ref{sec:discussion} explains.
Against BF16 weights on that BF16-only build, the MX build with MXFP4 weights still decodes 1.23$\times$-1.85$\times$ faster at 512, although it decodes Llama-3.2-1B at 16K 9\% slower.
Therefore, MXFP4 shortens only the weight streaming, and the fixed time that it cannot touch keeps the speedup below the byte ratio.

\input{1_Figures/TRETS/evaluation/fig-eval-gpu-float}

\textbf{Context length.}
Figure~\ref{fig-eval-board} also shows that a longer context lengthens every TPOT and shrinks the speedup of MXFP4 weights.
From 512 to 16K, the MXFP4 TPOT becomes 1.7$\times$ longer for Qwen3-1.7B and up to 2.5$\times$ longer for Llama-3.2-1B, because the platform rereads the KV cache of every earlier position for each query head.
The FP-FP mode multiplies these BF16 KV elements, so its proportion of the bytes of each token on the right axis rises from at most 72\% at 512 to at least 80\% at 16K.
The BF16 KV cache adds the same bytes to both formats, so both the byte ratio and the speedup decrease as the context grows.
For the three models measured in both formats at 16K, the average speedup decreases from 1.40$\times$ at 512 to 1.19$\times$ at 16K.
Consequently, a long context runs most of each token in the FP-FP mode, so a datapath must sustain this mode at the full HBM rate.

\textbf{Model size.}
The speedup of MXFP4 weights in Figure~\ref{fig-eval-board} also grows with the model size.
At a context of 512, the speedup rises from 1.38$\times$ for Llama-3.2-1B to 1.87$\times$ for Llama-3.1-8B, and from 1.26$\times$ for Qwen3-1.7B to 1.64$\times$ for Qwen3-8B.
Larger models keep more of their bytes in the MLP weights, so a BF16 token reads 2.19$\times$ the MXFP4 bytes on Llama-3.1-8B against 1.88$\times$ on Llama-3.2-1B.
The fixed time also grows more slowly than the weights, so its proportion of the MXFP4 TPOT at 512 drops from 57\% on Llama-3.2-1B to 27\% on Llama-3.1-8B.
Qwen3 models carry a larger fixed time than Llama models with the same layer count, because the VPU normalizes each of their query and key heads.
As a result, the MXFP4 gain grows with the model size, and the smaller weight footprint allows both 8B models to decode a 16K context that exceeds the HBM capacity in BF16.

\textbf{TTFT.}
Figure~\ref{fig-eval-board}(a) shows that MXFP4 weights shorten the TTFT even more than the TPOT.
The platform prefills a prompt token by token, so the TTFT follows the same byte argument as the TPOT.
Only the last prompt token runs the LM head and sampling, which raises the byte ratio of a prompt token and removes most of its fixed time.
Therefore, the TTFT speedup exceeds the TPOT speedup in every pair, with an average of 1.94$\times$ and a maximum of 2.19$\times$ at 512.
It decreases to an average of 1.39$\times$ at 16K, because later prompt tokens read the KV cache of more positions.

\textbf{Comparison with a GPU.}
Figure~\ref{fig-eval-gpu} shows that the A40 baseline of Section~\ref{sec-eval-setup} decodes faster than the board, whereas the board spends less energy per output token for every model and context except one case.
In single-batch decoding, the A40 decodes an average of 2.7$\times$ faster than the board at a context of 512, with a maximum of 3.6$\times$.
The A40 has only 1.5$\times$ the peak memory bandwidth of the board, so the rest of this gap comes from how much of its peak each platform uses.
For the 3B and 8B models at 512, the A40 streams the bytes of a token at 70\%-76\% of its 696~GB/s, whereas the fixed time holds the board at 45\%-55\% of its 460~GB/s.
The A40 also draws an average of 4.7$\times$ the card power of the board, so it spends an average of 1.75$\times$ the energy per output token of the board at 512.
Because Marlin lets the A40 read the same MXFP4 weight bytes, the energy advantage of the board comes from its lower power rather than from fewer bytes.
The speed gap widens to an average of 4.2$\times$ at 16K, because the A40 TPOT grows by only 12\%-66\% whereas the board rereads the KV cache for every query head.
Therefore, the energy ratio narrows to an average of 1.18$\times$ at 16K, and Llama-3.2-1B at 16K is the one case where the A40 spends less energy.
As a result, the board saves the most energy where the weights dominate the bytes of a token, which is also where MXFP4 weights help most.

\input{1_Figures/TRETS/evaluation/tab-eval-fpga}

\textbf{Comparison with FPGA accelerators.}
Table~\ref{tab-eval-fpga} places the board next to FlightLLM~\cite{zeng2024flightllm} and EdgeLLM~\cite{huang2025edgellm}, which run their own models and formats on their own boards, so the table only locates the platform among FPGA accelerators.
Both decode 1.4$\times$-1.5$\times$ faster than the board, because their sparse or INT4 weights take less than half the weight bytes per token.
Counted as weight bytes per token times tokens per second, the board uses 54\% of its peak bandwidth whereas the same count gives at most 35\% for both.
Therefore, the board delivers fewer tokens per second and per joule than both accelerators because it reads more bytes per token, whereas it uses more of its peak bandwidth.

Overall, both modes of the UDPU stream weights at about 93\% of the HBM port limit, so MXFP4 weights shorten the TPOT at a context of 512 by an average of 1.54$\times$ and by up to 1.87$\times$.
Any datapath that reaches this limit in both modes gains as much, whereas within the resources of 16 UDPU lanes no alternative of Section~\ref{sec-eval-alternatives} reaches it.

\subsection{Discussion}
\label{sec:discussion}

\textbf{Vector-unit bottleneck on GPT-OSS-20B.}
At a context of 512, the A40 of Section~\ref{sec-eval-e2e} decodes GPT-OSS-20B 3.0$\times$ faster than the board, against 2.1$\times$ for Llama-3.1-8B.
The gap is wider although GPT-OSS-20B reads fewer bytes per token, so it comes from work outside the dot-product units.
The RTL simulations of Section~\ref{sec-eval-setup} attribute 33\% of its token to sampling over 201,088 vocabulary entries and 13\% to vector operations, against 14\% and 7\% for Llama-3.1-8B.
Vector operations reach 26\% of its token at 8K, because the softmax with attention sinks of its full-attention layers grows with the context.

The platform adopted MXFP4 through a new bitstream, and its MX build halves the VPU to eight elements per cycle, because a full-width VPU beside the four UDPU lanes of SLR2 fails to route.
On the whole device, the MX build uses 2,328 DSP slices and 661K LUTs, whereas a BF16-only build of the same RTL uses 1,416 and 588K.
The MX build spends 64\% more DSP slices and 12\% more LUTs.
The vector operations of GPT-OSS-20B run on half the VPU width of the BF16-only build, whereas the A40 spreads sampling and vector operations over thousands of cores.
Thus, a wider VPU and faster sampling could narrow this gap without changing the UDPU, and the board already spends 1.2$\times$-1.8$\times$ less energy per token on this model.

\textbf{Beyond single-batch decoding.}
Every result of this evaluation holds for single-batch decoding, and batched serving changes two conditions behind the design of the UDPU.
The first is the sequential schedule, under which a split datapath idles one lane in every phase.
Iteration-level batching~\cite{yu2022orca} advances all requests through each layer together, so it keeps this schedule and the advantage of the UDPU.
Only a scheduler that partitions the batch into sub-batches in different layers~\cite{zhu2025nanoflow} runs the two modes concurrently.
Under such overlap, a split sized to the mode ratio needs fewer DSP slices than the UDPU, as Section~\ref{sec:split-unified} notes.
However, each sub-batch reads the weights of a layer again on a platform that streams weights from HBM, so the overlap trades weight reuse for concurrency.

The second is the proportion of time in the FP-FP mode, which batching raises because the KV traffic grows with the batch whereas the weight traffic does not.
Batching also makes the linear layers compute-bound in both modes, because the UDPU sizes both product rates to single-batch decoding.
The FP-FP mode then runs at only a quarter of the MX-FP rate, so batching lengthens this mode through both the KV traffic and its lower product rate.
Therefore, batched serving calls for higher product rates or an overlapped schedule.

%% file: 1_Figures/TRETS/evaluation/tab-eval-setup.tex
\begin{table}[t]
\centering
\caption{Configuration of the FPGA board and the GPU baseline. Both read the same MXFP4 weight bytes per token and keep the KV cache in BF16.}
\label{tab-eval-setup}
\footnotesize
\begin{tabular*}{\linewidth}{@{\extracolsep{\fill}}lll@{}}
\toprule
 & FPGA board & GPU baseline \\
\midrule
Device & AMD Alveo U55C (16~nm, 2021) & NVIDIA A40 (8~nm, 2020) \\
Memory & 16~GB HBM2 at 460~GB/s & 48~GB GDDR6 at 696~GB/s \\
Power limit & 150~W & 300~W \\
Clock & 185.4~MHz & 1,740~MHz \\
Host CPU & 2$\times$ Intel Xeon Silver 4310 (24 cores total) & 2$\times$ Intel Xeon Silver 4310 (24 cores total) \\
Host memory & 256~GB DDR4 & 256~GB DDR4 \\
Software & XRT 2.15 with in-house model compiler & vLLM 0.31.0, PyTorch 2.13, CUDA 13.0 \\
\bottomrule
\end{tabular*}
\end{table}

%% file: 1_Figures/TRETS/evaluation/tab-eval-variants.tex
\begin{table}[t]
\centering
\caption{Post-route resources per lane of the UDPU and its ablation variants. Parentheses give each $\Delta$ as a percentage of the UDPU lane, and DSP counts include eight accumulator slices.}
\label{tab-eval-variants}
\small
\begin{tabular*}{\linewidth}{@{\extracolsep{\fill}}llrrrrr@{}}
\toprule
Variant & Modes & LUT & FF & DSP & $\Delta$LUT & $\Delta$DSP \\
\midrule
\textbf{UDPU (ours)} & Both & 17,979 & 19,129 & 136 & 0 & 0 \\
MX-only, FP-FP mode removed & MX-FP & 15,302 & 17,741 & 136 & $-$2,677 ($-$14.9\%) & 0 \\
No packing, one product per slice & Both & 17,623 & 18,875 & 264 & $-$356 ($-$2.0\%) & $+$128 ($+$94.1\%) \\
No packing, LUT multipliers & Both & 28,554 & 28,480 & 8 & $+$10,575 ($+$58.8\%) & $-$128 ($-$94.1\%) \\
\bottomrule
\end{tabular*}
\end{table}

%% file: 1_Figures/TRETS/evaluation/tab-eval-stages.tex
\begin{table}[t]
\centering
\caption{Post-route resources of a UDPU lane and operand path by step. $\Delta$LUT is the change of the MX-only variant from the UDPU as a percentage of the step, and shared LUT sites lift column sums by up to 2.4\%.}
\label{tab-eval-stages}
\small
\def\halfdown#1{\smash{\raisebox{-0.6\normalbaselineskip}{#1}}}
\setlength{\aboverulesep}{0pt}\setlength{\belowrulesep}{0pt}\renewcommand{\arraystretch}{1.2}
\begin{tabular*}{\linewidth}{@{}l|l@{\hspace{2\tabcolsep}\extracolsep{\fill}}lrrrr@{}}
\toprule
Stage & Step & Modes & LUT & FF & DSP & $\Delta$LUT \\
\midrule
Operand & Activation pre-alignment & MX-FP & 1,136 (6.3\%) & 737 & 0 & 0 \\
path & BF16 activation buffering & FP-FP & 1,026 (5.7\%) & 1,282 & 0 & $-$1,024 ($-$99.8\%) \\
\midrule
\halfdown{Product} & Weight decoding and multipliers & Both & 6,032 (33.6\%) & 9,187 & 128 & $-$1,401 ($-$23.2\%) \\
 & Exponent adders & Both & 960 (5.3\%) & 1,152 & 0 & $-$128 ($-$13.3\%) \\
\midrule
\halfdown{Reduction} & Alignment & Both & 7,305 (40.6\%) & 4,343 & 0 & $-$215 ($-$2.9\%) \\
\halfdown{pipeline} & Summation & Both & 1,168 (6.5\%) & 2,136 & 0 & 0 \\
 & Normalization and accumulation & Both & 565 (3.1\%) & 293 & 8 & $+$28 ($+$5.0\%) \\
\midrule
Total & & & 17,979 (100\%) & 19,129 & 136 & $-$2,677 ($-$14.9\%) \\
\bottomrule
\end{tabular*}
\end{table}

%% file: 1_Figures/TRETS/evaluation/tab-eval-alternatives.tex
\begin{table}[t]
\centering
\caption{Post-route resources at the throughput of one UDPU lane with UDPU multiples, and products per cycle each design reaches within the resources of 16 UDPU lanes. BF16 lanes pack two products per DSP slice.}
\label{tab-eval-alternatives}
\def\namecell#1{\multicolumn{1}{@{}l@{\hspace{\tabcolsep}}|@{\hspace{\tabcolsep}}}{#1}}
\newlength{\namecolw}\settowidth{\namecolw}{\small FIGLUT-style}\addtolength{\namecolw}{\dimexpr\tabcolsep+\arrayrulewidth\relax}
\newlength{\subcolw}\settowidth{\subcolw}{\small MX-FP lane only}\addtolength{\subcolw}{2\tabcolsep}
\def\designsep{\noalign{\hbox{{\color{black!25}\vrule width\dimexpr\namecolw+\subcolw\relax height 0.4pt}}}}
\def\splitsep{\noalign{\hbox{\hskip\namecolw{\color{black!25}\vrule width\subcolw height 0.4pt}}}}
\def\tworow#1{\smash{\raisebox{-0.5\normalbaselineskip}{#1}}}
\small
\setlength{\aboverulesep}{0pt}\setlength{\belowrulesep}{0pt}\renewcommand{\arraystretch}{1.2}
\begin{tabular*}{\linewidth}{@{\extracolsep{\fill}}l@{\extracolsep{0pt}\hspace{2\tabcolsep}}l|l@{\ }l@{\extracolsep{\fill}\hspace{2\tabcolsep}}ll@{\extracolsep{0pt}\ }l|l@{\extracolsep{\fill}\hspace{2\tabcolsep}}l@{}}
\toprule
\multicolumn{2}{@{}l|}{Basis / Metric} & \multicolumn{5}{c|}{\begin{tabular}{@{}c@{}}Equal throughput /\\Resources\end{tabular}} & \multicolumn{2}{c}{\begin{tabular}{@{}c@{}}Equal resources /\\Products per cycle\end{tabular}} \\
\cmidrule{1-2} \cmidrule{3-7} \cmidrule{8-9}
\multicolumn{2}{@{}l|}{Design} & \multicolumn{2}{l}{LUT} & FF & \multicolumn{2}{l|}{DSP} & MX-FP & FP-FP \\
\midrule
\multicolumn{2}{@{}l|}{\textbf{UDPU (ours)}} & 17,979 & & 19,129 & 136 & & 4,096 & 1,024 \\
\designsep
\multicolumn{2}{@{}l|}{Bandwidth-matched split} & 27,946 & (1.55$\times$) & 26,215 & 176 & (1.29$\times$) & 2,635 & 659 \\
\designsep
\multicolumn{2}{@{}l|}{FP-only extension} & 57,182 & (3.18$\times$) & 34,092 & 160 & (1.18$\times$) & 1,288 & 1,288 \\
\cmidrule[\lightrulewidth]{1-2}
\namecell{\tworow{FIGNA-style}} & Both lanes & 44,058 & (2.45$\times$) & 21,211 & 183 & (1.35$\times$) & 1,671 & 418 \\
\splitsep
\namecell{} & MX-FP lane only & 32,121 & (1.79$\times$) & 12,517 & 143 & (1.05$\times$) & 2,293 & - \\
\cmidrule[\lightrulewidth]{1-2}
\namecell{\tworow{FIGLUT-style}} & Both lanes & 65,359 & (3.64$\times$) & 32,193 & 60 & (0.44$\times$) & 1,127 & 282 \\
\splitsep
\namecell{} & MX-FP lane only & 52,715 & (2.93$\times$) & 23,719 & 20 & (0.15$\times$) & 1,397 & - \\
\bottomrule
\end{tabular*}
\end{table}

%% file: 1_Figures/TRETS/evaluation/fig-eval-unit-error-float.tex
\begin{figure}[t]
\centering
\includegraphics[width=\linewidth]{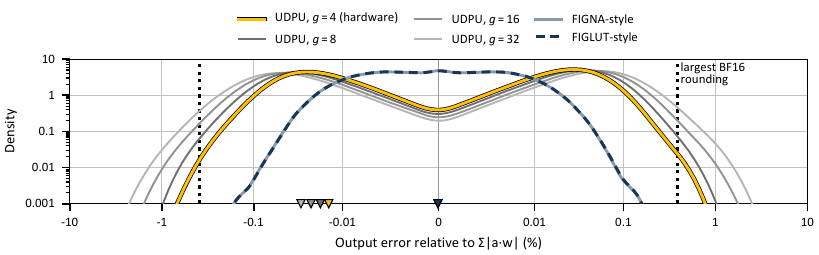}
\Description{Density curves of the output error on a symmetric log axis. The curves of the FIGNA-style and FIGLUT-style units lie within two dotted limit lines, and the curves of the UDPU are wider and widen further with the group size.}
\caption{Density of the RTL output error relative to $\sum |a w|$ on the MLP inputs of five dense models, with the UDPU at group sizes $g$. Dotted lines mark the largest BF16 rounding error, and triangles mark the mean error.}
\label{fig-eval-unit-error}
\end{figure}

%% file: 1_Figures/TRETS/evaluation/tab-eval-accuracy.tex
\begin{table}[t]
\centering
\caption{Accuracy with BF16 weights and with MXFP4 weights under exact products, the alternative alignments, or UDPU pre-alignment of group size $g$. Bold marks the hardware's $g = 4$, and $^{*}$ marks perplexity on UltraChat transcripts \mbox{($\uparrow$ means higher is better, $\downarrow$ means lower is better)}.}
\label{tab-eval-accuracy}
\footnotesize
\setlength{\tabcolsep}{3pt}
\begin{tabular*}{\linewidth}{@{\extracolsep{\fill}}llrrrrrrrrrr@{}}
\toprule
 & & \multicolumn{1}{c}{BF16 weights} & \multicolumn{9}{c}{MXFP4 weights} \\
\cmidrule(lr){3-3}\cmidrule(l){4-12}
 & & & & \multicolumn{2}{c}{Alternatives} & \multicolumn{6}{c}{UDPU, group size $g$} \\
\cmidrule(lr){5-6}\cmidrule(l){7-12}
Model & Metric & \multicolumn{1}{c}{Exact} & Exact & FIGNA & FIGLUT & $1$ & $2$ & $\mathbf{4}$ & $8$ & $16$ & $32$ \\
\midrule
Llama-3.2-1B & PPL$\downarrow$ & 15.01 & 16.51 & 16.51 & 16.51 & 16.54 & 16.55 & \textbf{16.56} & 16.57 & 16.59 & 16.62 \\
 & ARC-C$\uparrow$ & 41.0 & 39.2 & 39.2 & 38.7 & 39.2 & 39.7 & \textbf{39.4} & 39.5 & 39.5 & 39.2 \\
 & MMLU$\uparrow$ & 47.0 & 45.1 & 45.1 & 45.6 & 45.3 & 45.5 & \textbf{45.4} & 45.4 & 44.6 & 45.2 \\
\midrule
Llama-3.2-3B & PPL$\downarrow$ & 12.57 & 13.36 & 13.36 & 13.36 & 13.40 & 13.40 & \textbf{13.41} & 13.42 & 13.45 & 13.46 \\
 & ARC-C$\uparrow$ & 50.7 & 49.4 & 49.6 & 49.5 & 49.2 & 49.3 & \textbf{49.7} & 49.8 & 49.9 & 49.3 \\
 & MMLU$\uparrow$ & 60.8 & 60.5 & 60.2 & 60.5 & 60.4 & 60.4 & \textbf{60.9} & 60.4 & 60.3 & 60.4 \\
\midrule
Llama-3.1-8B & PPL$\downarrow$ & 8.05 & 8.50 & 8.50 & 8.50 & 8.51 & 8.51 & \textbf{8.52} & 8.52 & 8.53 & 8.54 \\
 & ARC-C$\uparrow$ & 60.9 & 59.8 & 59.8 & 59.9 & 59.5 & 59.9 & \textbf{59.7} & 59.6 & 60.1 & 59.4 \\
 & MMLU$\uparrow$ & 68.6 & 66.8 & 66.8 & 67.0 & 67.0 & 67.1 & \textbf{67.1} & 67.0 & 66.9 & 66.7 \\
\midrule
Qwen3-1.7B & PPL$\downarrow$ & 18.55 & 18.78 & 18.78 & 18.78 & 18.89 & 18.90 & \textbf{18.91} & 18.92 & 18.97 & 19.14 \\
 & ARC-C$\uparrow$ & 52.3 & 50.3 & 50.3 & 50.8 & 50.9 & 50.5 & \textbf{50.9} & 50.7 & 50.0 & 50.0 \\
 & MMLU$\uparrow$ & 62.6 & 59.9 & 59.9 & 60.1 & 59.9 & 59.8 & \textbf{59.9} & 59.9 & 59.6 & 59.6 \\
\midrule
Qwen3-8B & PPL$\downarrow$ & 10.94 & 11.25 & 11.25 & 11.25 & 11.28 & 11.28 & \textbf{11.28} & 11.28 & 11.28 & 11.29 \\
 & ARC-C$\uparrow$ & 66.3 & 65.5 & 65.5 & 65.3 & 65.3 & 65.3 & \textbf{65.2} & 65.1 & 65.0 & 65.4 \\
 & MMLU$\uparrow$ & 77.2 & 76.5 & 76.4 & 76.6 & 76.5 & 76.5 & \textbf{76.6} & 76.3 & 76.4 & 76.3 \\
\midrule
GPT-OSS-20B & PPL$^{*}\downarrow$ & -- & 3.39 & 3.39 & 3.38 & 3.38 & 3.37 & \textbf{3.37} & 3.35 & 3.35 & 3.35 \\
 & ARC-C$\uparrow$ & -- & 92.8 & 93.8 & 93.0 & 93.9 & 93.9 & \textbf{94.1} & 93.8 & 93.4 & 93.8 \\
 & MMLU$\uparrow$ & -- & 83.2 & 83.3 & 83.1 & 83.2 & 82.8 & \textbf{82.8} & 83.1 & 83.4 & 83.3 \\
\bottomrule
\end{tabular*}
\end{table}

%% file: 1_Figures/TRETS/evaluation/fig-eval-board-float.tex
\begin{figure}[t]
\centering
\includegraphics[width=\linewidth]{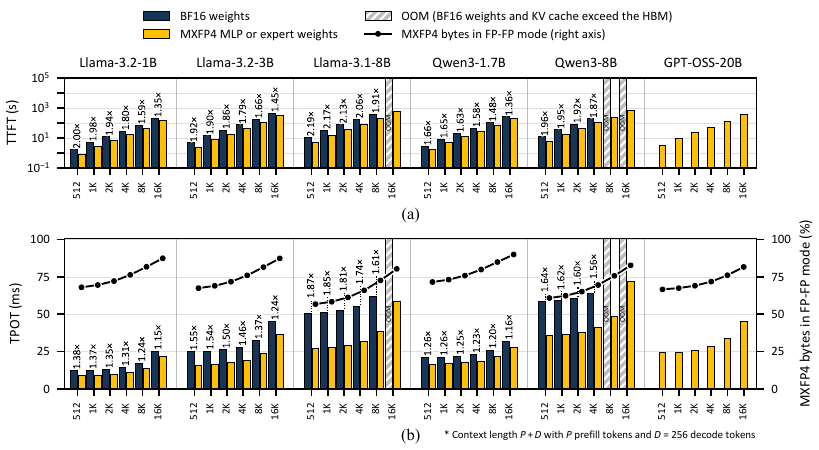}
\Description{Bar charts of the time to first token and the time per output token on the board for six models at six context lengths with BF16 and MXFP4 weights.}
\caption{End-to-end decoding on the U55C board as the context grows. (a) TTFT on a log scale and (b) TPOT with the MXFP4 speedup above each pair, where the right axis of (b) gives the proportion of FP-FP bytes.}
\label{fig-eval-board}
\end{figure}

%% file: 1_Figures/TRETS/evaluation/fig-eval-gpu-float.tex
\begin{figure}[t]
\centering
\includegraphics[width=\linewidth]{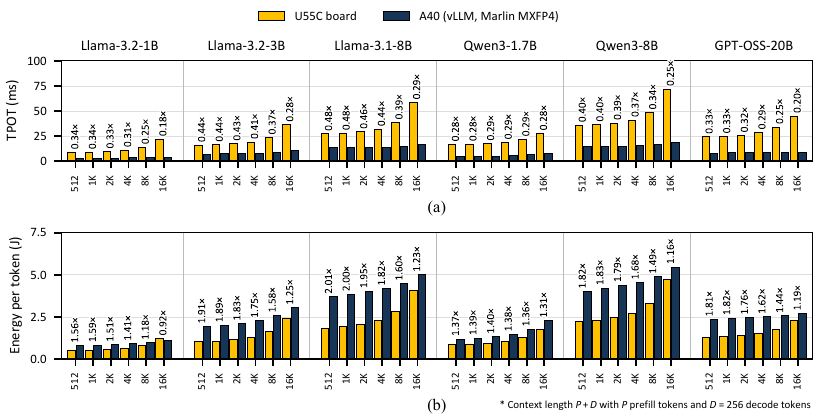}
\Description{Bar charts of the time per output token and the energy per output token of the board and of an A40 GPU for five models at six context lengths.}
\caption{Single-batch MXFP4 decoding on the U55C board and on an A40 GPU with vLLM and Marlin kernels. (a) TPOT and (b) energy per output token, with the A40 result over the board result above each pair.}
\label{fig-eval-gpu}
\end{figure}

%% file: 1_Figures/TRETS/evaluation/tab-eval-fpga.tex
\begin{table}[t]
\centering
\caption{\sysname next to FPGA accelerators that report single-batch decoding of a 6B-8B model on one 460~GB/s HBM board. $\approx$ marks values read from figures, and $\leq$ a dense byte count of sparse weights.}
\label{tab-eval-fpga}
\footnotesize
\setlength{\tabcolsep}{3pt}
\begin{tabular*}{\linewidth}{@{\extracolsep{\fill}}llllrrrrrr@{}}
\toprule
Design & Board & Model & Weights / act. & GB/token & tok/s & BW util. & W & tok/J & LUT / DSP \\
\midrule
FlightLLM~\cite{zeng2024flightllm} & U280 & LLaMA2-7B & 3.5-bit sparse / INT8 & $\leq$2.95 & $\approx$55 & $\leq$35\% & $\approx$45 & $\approx$1.22 & 44\% / 70\% \\
EdgeLLM~\cite{huang2025edgellm} & VCU128 & ChatGLM2-6B & INT4 / FP16 & 3.08 & 52.7 & 35\% & 56.9 & 0.93 & 74\% / 51\% \\
\textbf{\sysname (ours)} & U55C & Llama-3.1-8B & MXFP4 MLP / BF16 & 6.73 & 36.8 & 54\% & 67.4 & 0.55 & 33\% / 26\% \\
\bottomrule
\end{tabular*}
\end{table}

%% file: 0_Text/8_related_work.tex
\section{Related Work}
\label{sec:related}

\textbf{Arithmetic units for quantized LLMs.}
FP-INT units avoid a dequantizer by multiplying quantized weights with pre-aligned integer arithmetic, partial-product tables, or approximate multipliers~\cite{jang2024figna,park2025figlut,mo2025lut,zou2025axcore}.
These units leave the BF16$\times$BF16 products of attention to the separate FP unit that Section~\ref{sec-back_dpu} describes, unless the KV cache is quantized as well~\cite{zou2025axcore}.
In contrast, OPAL and MicroScopiQ quantize the activations to block-scaled integers~\cite{koo2024opal,ramachandran2025microscopiq}, which lets OPAL compute $QK^\top$ on its integer units.
As Section~\ref{sec-arch-prealign} describes, \sysname takes BF16 activations and aligns them online in groups of four only inside the MX-FP mode.
It keeps the attention products in BF16 for the reasons of Section~\ref{sec:precision} and executes both modes in one datapath rather than two units.

\textbf{Microscaling hardware.}
MSFP brought shared-exponent arithmetic to production on the FPGAs of Project Brainwave~\cite{rouhani2020msfp}, and shared microexponents refined it into the OCP MX formats~\cite{darvish2023shared,ocp_mx_spec_v1}.
Blackwell tensor cores now execute MX formats natively~\cite{nvidia_blackwell_datasheet}.
On FPGAs, Samson et al.~\cite{10705505} implement the dot products of the MX standard in an open-source library, and Abdurakhmanov and Fahmy~\cite{abdurakhmanov2025mx} map systolic arrays for 6-bit and 8-bit MX formats onto UltraScale+ devices.
All of these designs scale every operand by block, whereas the MX-FP mode pairs an MX weight block with BF16 activations whose exponents differ within the block.
Therefore, the UDPU applies the shared scale apart from the multipliers and aligns the products once per group, as Sections~\ref{sec-arch-scale} and~\ref{sec-arch-prealign} describe.

\textbf{FPGA accelerators for LLM inference.}
LLM accelerators on FPGAs range from multi-board FP16 appliances to designs with sparse or low-bit integer weights~\cite{hong2022dfx,moon2024lpu,zeng2024flightllm,huang2025edgellm,chen2024spatial,ma2026cdllm}.
Each of them builds its datapath around the weight formats it targets, and none multiplies MX weights by BF16 activations.
The UDPU brings this product to a deployed board through a new bitstream, whereas fixed silicon of its generation reaches the format only through software dequantization, as Section~\ref{sec-eval-setup} describes.

\textbf{DSP packing on FPGAs.}
Packing several narrow products into one DSP multiplier is an established way to raise the products per slice, from the INT8 scheme of AMD to INT4$\times$INT8 and low-bit convolution packing~\cite{fu2017wp486,chen2024spatial,liu2022hikonv,yang2024acane}.
DSP-Packing generalizes it to signed integers of any width and corrects the packed results with fabric adders or approximately through the C port~\cite{sommer2022dsppacking}.
All of these schemes multiply integer operands, whereas the UDPU of Section~\ref{sec-arch-dsp} packs two signed MXFP4$\times$BF16 mantissa products that one C-port term keeps exact and bit-slices BF16$\times$BF16 products onto the same slices.

\textbf{Single datapaths for both modes.}
Composable and reconfigurable units serve several formats with one datapath~\cite{sharma2018bitfusion,wang2026harmonia,zhou2026mxglut,wu2025tataa}, but they quantize the activations of every matrix multiplication, so none executes the BF16$\times$BF16 products of attention.
XtraMAC and SkipOPU come closest to \sysname, because they run low-precision and FP products on the same DSP slices of an FPGA~\cite{yu2026xtramac,he2026skipopu}.
XtraMAC offsets the exponent of every product and SkipOPU quantizes weights per channel, so neither applies the MX shared scale once per group.
The FP-only extension of Section~\ref{sec-eval-alternatives} aligns every product as XtraMAC does and reaches 3.2$\times$ the LUTs of the UDPU at the MX-FP rate.

%% file: 0_Text/9_conclusion.tex
\section{Conclusion}
\label{sec:conclu}

In this paper, we present \sysname, whose UDPU executes the MX-FP and FP-FP modes of MXFP4 inference on one datapath.
We propose per-group pre-alignment to share one reduction pipeline and pre-adder packing to share one DSP array between the two modes.
Against four alternative datapaths on the same FPGA, the UDPU reaches the same per-mode throughput with 36\%-72\% fewer LUTs.
It also needs 15\%-26\% fewer DSP slices than every alternative that multiplies in DSP slices, and its pre-alignment changes perplexity by at most 0.7\%.
On an AMD Alveo U55C board, MXFP4 weights shorten the TPOT by 1.15$\times$-1.87$\times$.
An NVIDIA A40 GPU decodes faster but spends 0.92$\times$-2.01$\times$ the energy per token of the board.
However, we scope these results to single-batch decoding with a BF16 KV cache, so an MX-FP mode for MXFP8 keys and values under batched serving remains future work.
We believe that unified datapaths let FPGAs adopt the number formats of LLMs as they emerge.